\documentclass[aps,prb,showpacs,floatfix,preprint]{revtex4}

\usepackage{graphicx}
\usepackage{amssymb}

\begin{document}

\markboth{A. Sherman}
{Magnetic incommensurability in $p$-type cuprate perovskites}

\title{Magnetic incommensurability in $p$-type cuprate perovskites}

\author{A. Sherman}

\affiliation{Institute of Physics, University of Tartu, Riia 142,
 51014 Tartu, Estonia}

\date{\today}

\begin{abstract}
For the superconducting phase with a $d$-wave order parameter and zero temperature the magnetic susceptibility of the $t$-$J$ model is calculated using the Mori projection operator technique. Conditions for the appearance of an incommensurate magnetic response below the resonance frequency are identified. A fast decay of the tails of the hole coherent peaks and a weak intensity of the hole incoherent continuum near the Fermi level are enough to produce an incommensurate response using different hole dispersions established for $p$-type cuprates, in which such response was observed. In this case, the nesting of the itinerant-electron theory or the charge modulation of the stripe theory is unnecessary for the incommensurability. The theory reproduces the hourglass dispersion of the susceptibility maxima with their location in the momentum space similar to that observed experimentally. The upper branch of the dispersion stems from the excitations of localized spins, while the lower one is due to the incommensurate maxima of their damping. The narrow and intensive resonance peak arises if the frequency of these excitations at the antiferromagnetic momentum lies below the edge of the two-fermion continuum; otherwise the maximum is broad and less intensive.
\end{abstract}


\maketitle

\section{Introduction}
One of the most interesting features of the inelastic neutron scattering in lanthanum cuprates is that for hole concentrations $x\gtrsim 0.04$, low temperatures and small energy transfers the scattering intensity is peaked at the incommensurate momenta $\left(\frac{1}{2},\frac{1}{2}\pm\kappa\right)$, $\left(\frac{1}{2}\pm\kappa, \frac{1}{2}\right)$ in the reciprocal lattice units $2\pi/a$ with the lattice period $a$.\cite{Yoshizawa,Birgeneau} For $x\lesssim 0.12$ the incommensurability parameter $\kappa$ is approximately equal to $x$ and saturates for larger concentrations.\cite{Yamada} The incommensurability was observed both below and above the temperature of the superconducting transition $T_c$.\cite{Mason,Matsuda} The analogous low-frequency incommensurability was observed also in YBa$_{2}$Cu$_{3}$O$_{7-y}$ and Bi$_{2}$Sr$_{2}$CaCu$_{2}$O$_{8}$.\cite{Dai,Arai,Bourges} This gives ground to suppose that the incommensurability is a common feature of $p$-type cuprate perovskites. As the transfer frequency increases, the incommensurability parameter $\kappa$ decreases and at the frequency $\omega _r\approx 25-40$~meV the magnetic response becomes commensurate, being peaked at the antiferromagnetic wave vector ${\bf Q}=\left(\frac{1}{2},\frac{1}{2}\right)$.\cite{Bourges98,He} The value of the $\omega_r$ depends on the hole concentration. For even larger frequencies in some crystals the magnetic response becomes again incommensurate, with the incommensurability parameter growing
with frequency.\cite{Bourges98,He} In some experiments susceptibility maxima were found at momenta $\left(\frac{1}{2}\pm\kappa,\frac{1}{2}\pm\kappa\right)$, $\left(\frac{1}{2}\pm\kappa,\frac{1}{2}\mp\kappa\right)$.\cite{Bourges98,He,Haiden,Tranquada} In other experiments the maxima form a circle around ${\bf Q}$ or even merge together into a broad commensurate maximum.\cite{Xu,Endoh,Stock} Thus, the dispersion of the maxima in the susceptibility resembles a hourglass with the neck at the momentum ${\bf Q}$ and the frequency
$\omega_r$.\cite{Dai,Arai,Bourges,Tranquada} In YBa$_{2}$Cu$_{3}$O$_{7-y}$, in the region of the neck the susceptibility peak is much more intensive and narrower than the maxima for smaller and larger frequencies. This peculiarity of the spin-excitation spectrum was called the resonance peak.\cite{Rossat}

The most frequently cited theories for the magnetic incommensurability are based on the picture of itinerant electrons with the magnetic susceptibility calculated in the random phase approximation\cite{Lavagna,Liu,Bulut,Brinckmann,Norman} and on the stripe picture.\cite{Tranquada,Uhrig,Vojta,Seibold} In the former approach the  incommensurability below $\omega_r$ is connected with a nesting in the constant energy surfaces of the carrier dispersion. The approach allows one to reproduce the observed hourglass dispersion of the susceptibility maxima. However, carrier dispersions derived from photoemission experiments frequently have no nesting,\cite{Ino} and some fitting of their parameters is necessary to obtain the incommensurability.\cite{Norman} Such sensitivity of the results to fine details of the hole dispersion casts some doubt upon this approach, since in different $p$-type cuprates, in which the incommensurability is observed, the dispersions diverge considerably and vary with doping. Another theory\cite{Eremin05} based on the itinerant-electron approximation relates the incommensurability to a singularity in the noninteracting susceptibility, which arises on the assumption of vanishingly small damping of fermion states.\cite{Brinckmann,Abanov}

The second approach is based on the notion of charge stripes -- a one-dimensional periodic variation of the charge density in a Cu-O plane. This theory also reproduces the observed hourglass dispersion of the susceptibility maxima with their proper location in the Brillouin zone, at least for $x=\frac{1}{8}$. It should be noted that static charge stripes are observed only in crystals with the low-temperature tetragonal (LTT) or the less-orthorhombic phases (La$_{2-y-x}$Nd$_{y}$Sr$_{x}$CuO$_{4}$ and La$_{2-x}$Ba$_{x}$CuO$_{4}$), and such stripes are not detected in the low-temperature orthorhombic (LTO) phase of cuprates, e.g., in La$_{2-x}$Sr$_{x}$CuO$_{4}$.\cite{Kimura,Fujita} At the same time the magnetic incommensurability is similar in these phases. To resolve this contradiction the supposition of stripe fluctuations in the LTO phase with the same symmetry as in the LTT phase was proposed.\cite{Vojta} However, there are strong grounds to believe that stripes are connected with deformations of Cu-O planes due to certain tilts of CuO$_6$ octahedra.\cite{Barisic,Pickett,Sherman08} In this case in the LTO phase dominant fluctuations will be oriented along diagonals of the planes.\cite{Kimura00} With such fluctuations the susceptibility maxima below $\omega_r$ will be located along diagonals of the Brillouin zone, in contrast to experimental observations. Besides, at frequencies $\omega >\omega _r$ the stripe theory predicts maxima of the magnetic susceptibility, which are sharply peaked in the momentum space at $\left(\frac{1}{2}\pm\kappa,\frac{1}{2}\pm\kappa\right)$, $\left(\frac{1}{2}\pm\kappa, \frac{1}{2}\mp\kappa\right)$. In experiments the susceptibility looks more isotropic in this region.\cite{Tranquada,Xu}

Strong electron correlations in Cu-O planes were taken into account in Refs.~\onlinecite{Sherman04,Sega} in calculating the magnetic susceptibility of the two-dimensional (2D) $t$-$J$ model. To carry out calculations with the Hubbard operators entering into the Hamiltonian the Mori projection operator technique\cite{Mori} was used. In Ref.~\onlinecite{Sherman04}, attention was drawn to the fact that for small frequencies the incommensurate response points to a dip in the spin-excitation damping at ${\bf Q}$. The dip was related to the nesting of small pockets around $\left(\pm\frac{1}{4},\pm\frac{1}{4}\right)$, which form the Fermi surface of the $t$-$J$ model at small $x$. The nesting is a consequence of the short-range antiferromagnetic order with a large correlation length, which is established for small hole concentrations. Some similarity between equations for the susceptibility obtained in the Mori and itinerant-electron approaches was pointed out in Ref.~\onlinecite{Sega}. In this work, the arguments of the itinerant-electron picture were used for the explanation of the incommensurability below and above $\omega_r$. In the case of strong electron correlations the incommensurability above the resonance frequency was related to the dispersion of excitations of localized spins in Refs.~\onlinecite{Barzykin,Sherman03a}. In Ref.~\onlinecite{Vladimirov} this part of the hourglass dispersion was also obtained in the approach based on the Mori projection operator technique. However, due to a poor representation for the operator, which determines the self-energy of spin excitations, the contribution of the decay into an electron-hole pair was lost. Together with it the ability to describe the low-frequency incommensurability was gone.

In the present work, we calculate the magnetic susceptibility of the 2D $t$-$J$ model using the Mori projection operator technique. We take into account all contributions to the polarization operator of spin excitations. These contributions are connected with a decay of a spin excitation into a fermion pair, a decay into the pair assisted by a hole or with a creation of another spin excitation, and a decay into three spin excitations. In the calculations, we take into account an incoherent hole-spin-excitation (HSE) continuum, which holds the major part of the quasiparticle weight of the hole spectral function in the considered case of strong correlations. For the coherent part of the spectral function several dispersions found in the literature were analyzed, including those designed to fit photoemission data. For the superconducting state with the $d$-wave order parameter $\Delta_{\bf k}$ we found that at certain conditions all considered dispersions produce an incommensurate magnetic response below the resonance frequency. This differs from the result obtained in the itinerant-electron picture where, as mentioned above, a special fitting of the hole dispersion was necessary to obtain the incommensurability. The cause of the difference is in the following. As it will be seen below, for moderate hole concentrations and frequencies less than 150~meV, there is a certain similarity in the description of the magnetic susceptibility in the Mori and itinerant-electron approaches. In the latter theory, $\delta$-functions in the imaginary part of the noninteracting susceptibility are usually approximated by Lorentzians with an artificial broadening.\cite{Bulut,Norman} An inspection of this approximation shows that far tails of the Lorentzians strengthen the intensity at ${\bf Q}$. In this situation, the Fermi surface has to contain nearly parallel regions for the intensity at the respective incommensurate wave vector be larger than at the commensurate ${\bf Q}$. The matter concerns tails, which are separated from the Lorentzian maximum by frequencies of the order or larger than the superconducting gap $\Delta=\max(\Delta_{\bf k})$, i.e. more than an order of magnitude larger than estimated widths of spectral peaks. Such long tails imply that decay widths of states near the Fermi level are independent of frequency. However, as known, the damping grows with distance from the Fermi level.\cite{Abrikosov,Varma} Therefore, the tails of the peaks decay much faster than in the Lorentzian. Besides, the tails have to be cut off in the superconducting gap. To take these facts into account, we approximated the $\delta$-functions, which ensure the energy conservation, by a $\Pi$-shaped step function of a finite width. This approximation gives the incommensurate magnetic response below the resonance frequency for the mentioned dispersions provided that the intensity of the HSE continuum is weak in the ranges $|\Delta_{\bf k}|<|\omega|<\Delta$.

This approach allows us to describe the hourglass dispersion of the susceptibility maxima, in which the part above $\omega_r$ reflects the dispersion of spin excitations. The maxima below $\omega_r$ originate from incommensurate peaks of the spin-excitation damping. They are sharply peaked at the axes of the Brillouin zone, while above the resonance frequency the susceptibility is more isotropic. Notice, however, that for one of the considered hole dispersions the low-frequency maxima were located on the diagonals, and in some cases the high-frequency incommensurate peaks merge into a broad commensurate maximum. The evolution of the Fermi surface and the increase of the spin-excitation frequency $\omega_{\bf Q}$ with doping explains rather naturally the observed growth of the incommensurability parameter $\kappa$ with $x$. In this approach, a narrow and intensive resonance peak arises in the susceptibility when the frequency $\omega_{\bf Q}$ falls on the region of small damping below the edge of the two-fermion states at the antiferromagnetic wave vector. If $\omega_{\bf Q}$ exceeds this threshold, a broader and less intensive maximum appears.

The main formulas used in the calculations are given in the next section. In more details their derivation is presented in the Appendix. The discussion of the obtained results is carried out in Sec.~III. Section IV is devoted to the speculation about possible changes in the susceptibility in the pseudogap and phase separated states and to concluding remarks.

\section{Main formulas}
The Hamiltonian of the 2D $t$-$J$ model reads
\begin{equation}\label{Hamiltonian}
H=\sum_{{\bf ll'}\sigma}t_{\bf ll'}a^\dagger_{{\bf l}\sigma}a_{{\bf l'}\sigma}+\frac{1}{2} \sum_{\bf ll'} J_{\bf ll'}\left(s^z_{\bf l}s^z_{\bf l'}+s^{+1}_{\bf l}s^{-1}_{\bf l'}\right)-\mu\sum_{\bf l} X^{00}_{\bf l},
\end{equation}
where $a_{{\bf l}\sigma}=|{\bf l}\sigma\rangle\langle{\bf l}0|$ is the hole annihilation operator, $|{\bf l}\sigma\rangle$ and $|{\bf l}0\rangle$ are singly occupied and empty states on a site ${\bf l}$ of a 2D square lattice, which models a Cu-O plane of $p$-type cuprate perovskites, $\sigma=\pm 1$, $t_{\bf ll'}$ and $J_{\bf ll'}$ are hopping and exchange constants, $s^z_{\bf l}=\frac{1}{2}\sum_\sigma \sigma|{\bf l}\sigma\rangle \langle{\bf l}\sigma|$ and $s^\sigma_{\bf l}=|{\bf l}\sigma \rangle\langle{\bf l}, -\sigma|$ are components of the spin-$\frac{1}{2}$ operator, the Hubbard operator $X^{00}_{\bf l}= |{\bf l}0 \rangle\langle{\bf l}0|$, and $\mu$ the chemical potential.

Our aim is the calculation of Green's function of spin excitations $D({\bf k}\omega)$, which is connected with the magnetic susceptibility by the relation\cite{Bourges98}
\begin{equation}\label{susceptibility}
\chi({\bf k}\omega)=-4\mu^2_B D({\bf k}\omega)
\end{equation}
with the Bohr magneton $\mu_B$. The function reads
\begin{equation}\label{Green}
D({\bf k}\omega)=\langle\langle s^z_{\bf k}|s^z_{\bf -k}\rangle\rangle =\int^\infty_{-\infty}e^{i\omega t}D({\bf k}t)d t,\quad
D({\bf k}t)=-i\theta(t)\left\langle\left[s^z_{\bf k}(t),s^z_{\bf -k}\right]\right \rangle.
\end{equation}
In this equation, the averaging denoted by the angle brackets and the time dependence of the operator are defined by Hamiltonian~(\ref{Hamiltonian}), $s^z_{\bf k}=N^{-1/2}\sum_{\bf l}e^{-i{\bf kl}}s^z_{\bf l}$, and $N$ is the number of lattice sites.

In the considered case of strong electron correlations it is convenient to use the Mori projection operator formalism\cite{Mori} for calculating Green's function (\ref{Green}). The respective formulas are given in the Appendix. The obtained expression for the function reads
\begin{equation}\label{GfD}
D({\bf k}\omega)=\frac{h_{\bf k}}{\omega^2-\omega\Pi({\bf k}\omega)-\omega_{\bf k}^2}.
\end{equation}
In this equation,
\begin{equation}\label{a1a1}
h_{\bf k}=4\left(tF_1+JC_1\right)\left(\gamma_{\bf k}-1\right),
\end{equation}
\begin{equation}\label{wk}
\omega^2_{\bf k}=16J^2\alpha|C_1|\left(1-\frac{tF_1}{J\alpha|C_1|} \right)\left(1-\gamma_{\bf k}\right)\left(\delta+1+\gamma_{\bf k}\right),
\end{equation}
\begin{equation}\label{Delta}
\delta=\left(1-\frac{tF_1}{J\alpha|C_1|}\right)^{-1}\bigg(
\frac{2t^2F_2}{J^2\alpha|C_1|}-\frac{tF_1}{4J\alpha|C_1|}+\frac{1}{4} +\frac{C_2}{|C_1|}+\frac{(1-\alpha)(1-x)}{8\alpha|C_1|}\bigg)-1,
\end{equation}
where we took into account only the nearest neighbor hopping and exchange terms in Hamiltonian (\ref{Hamiltonian}), which are characterized by the constants $t$ and $J$, respectively. The multiplier $\alpha$ is introduced to correct the decoupling procedure\cite{Kondo,Shimahara} (see the Appendix),
$$C_n=\frac{1}{N}\sum_{\bf k}\gamma^n_{\bf k}\langle s^{+1}_{\bf k}s^{-1}_{\bf k}\rangle,\quad F_n=\frac{1}{N}\sum_{\bf k}\gamma^n_{\bf k}\langle a^\dagger_{\bf k\sigma}a_{\bf k\sigma}\rangle,$$
$a_{\bf k\sigma}=N^{-1/2}\sum_{\bf l}e^{i{\bf kl}}a_{\bf l\sigma}$, $s^\sigma_{\bf k}=N^{-1/2}\sum_{\bf l}e^{-i\sigma{\bf  kl}}s^\sigma_{\bf l}$ and $\gamma_{\bf k}=\frac{1}{2}\left[\cos(k_x)+\cos(k_y)\right]$. Here and below the lattice spacing $a$ is set as the unit of length.

In Eq.~(\ref{GfD}), the polarization operator is given by the expression
\begin{eqnarray}\label{Pi}
\Pi({\bf k}\omega)&=&-h^{-1}_{\bf k}\Bigg\{\frac{2}{N^2}\sum_{\bf qq'} \left[2f_1^2({\bf kqq'})+f_3^2({\bf  kqq'})\right]\int\!\!\!\int\!\!\!\int_{-\infty}^\infty d\omega_1d\omega_2d\omega_3\nonumber\\
&&\quad\quad\times\frac{N_{BF}(\omega_1\omega_2\omega_3)B({\bf k'}\omega_1) \left[A({\bf  q}\omega_2)A({\bf  q'}\omega_3)-A'({\bf  q}\omega_2)A'({\bf  q'}\omega_3)\right]}{(\omega_2-\omega_1-\omega_3)(\omega+\omega_2-\omega_1-\omega_3 +i\eta)}\nonumber\\
&&\quad +\frac{4}{N^2}\sum_{\bf  qq'}f_4^2({\bf  kqq'})\int\!\!\!\int\!\!\!\int_{-\infty}^\infty d\omega_1d\omega_2d\omega_3 \nonumber\\ &&\quad\quad\times\frac{N_B(\omega_1\omega_2\omega_3)B({\bf  k'}\omega_1)B({\bf  q}\omega_2)B({\bf  q'}\omega_3)}{(\omega_2-\omega_1-\omega_3) (\omega+\omega_2-\omega_1-\omega_3 +i\eta)}\nonumber\\
&&\quad+\frac{2}{N}\sum_{\bf  q}f_2^2({\bf  kq})\int\!\!\!\int_{-\infty}^\infty d\omega_1d\omega_2\nonumber\\
&&\quad\quad\times\frac{N_F(\omega_1\omega_2)\left[A({\bf  k+q},\omega_1)A({\bf  q}\omega_2)+A'({\bf  k+q},\omega_1)A'({\bf  q}\omega_2)\right]}{(\omega_1-\omega_2)(\omega+\omega_1-\omega_2+i\eta)}\Bigg\},
\end{eqnarray}
where
\begin{eqnarray}
f_1({\bf  kqq'})&=&\frac{1}{2}\bigg[\varphi_1({\bf  q-k,k'})-\varphi_1({\bf  -q'-k,k'})+\frac{1}{2}\varphi_2({\bf  -q'-k,q})
+\varphi_2({\bf  -q',q})\nonumber\\
&&\quad -\frac{3}{2}\varphi_2({\bf  -q',q-k})+\frac{3}{2}\varphi_2({\bf  -q,q'+k})
-\frac{1}{2}\varphi_2({\bf  k-q,q'})\nonumber\\
&&\quad -\varphi_2({\bf  -q,q'})\bigg], \nonumber\\
f_2({\bf  kq})&=&\frac{1+x}{2}\bigg[\varphi_1({\bf  -q-k,k})-\frac{1}{2}\varphi_1({\bf  -q-k},{\bf 0})-\frac{1}{2}\varphi_1({\bf  q},{\bf 0})\bigg],\label{fi}\\
f_3({\bf  kqq'})&=&\varphi_1({\bf  -q,q-q'})-\frac{1}{2}\varphi_1({\bf  -q'-k,k'})
-\frac{1}{2}\varphi_1({\bf  q-k,k'})+\frac{1}{4}\varphi_2({\bf  k-q,q'})\nonumber\\
&&+\frac{1}{4}\varphi_2({\bf  -k-q',q})-\frac{1}{4}\varphi_2({\bf  -q',q-k})-\frac{1}{4}\varphi_2({\bf  -q,q'+k}),\nonumber\\
f_4({\bf  kqq'})&=&\varphi_3({\bf  q,q'-q})-\frac{1}{2}\varphi_3({\bf  -q,k+q'})
-\frac{1}{2}\varphi_3({\bf  q',k-q})-\frac{1}{2}\varphi_3({\bf  q-k,k'})\nonumber\\
&&+\frac{1}{2}\varphi_3({\bf  -k-q',q})+\frac{1}{2}\varphi_3({\bf  q-k,-q'})-\frac{1}{2}\varphi_3({\bf  -k-q',k'}),\nonumber
\end{eqnarray}
${\bf  k'=k-q+q'}$, $\eta=+0$,
\begin{eqnarray}
\varphi_1({\bf  qq'})&=&\sum_{\bf  p}\left(\delta_{\bf  pq}-\frac{1}{N}\right)t_{\bf  p}t_{\bf  p+q'}
=t_{\bf  q}t_{\bf  q+q'}-4t^2\gamma_{\bf  q'}-4(t')^2\gamma'_{\bf  q'},\nonumber\\
\varphi_2({\bf  qq'})&=&\sum_{\bf  p}\left(\delta_{\bf  pq}-\frac{1}{N}\right)t_{\bf  p}J_{\bf  p+q'}
=t_{\bf  q}J_{\bf  q+q'}-4tJ\gamma_{\bf  q'},\label{phi}\\
\varphi_3({\bf  qq'})&=&\sum_{\bf  p}\left(\delta_{\bf  pq}-\frac{1}{N}\right)J_{\bf  p}J_{\bf  p+q'}
=J_{\bf  q}J_{\bf  q+q'}-4J^2\gamma_{\bf  q'}\nonumber,
\end{eqnarray}
\begin{eqnarray}
N_{BF}(\omega_1\omega_2\omega_3)&=&\left[1+n_B(\omega_1)\right]n_F(\omega_2)\left[1 -n_F(\omega_3)\right]
-n_B(\omega_1)\left[1-n_F(\omega_2)\right]n_F(\omega_3),\nonumber\\
N_F(\omega_1\omega_2)&=&n_F(\omega_1)-n_F(\omega_2),\label{NBF}\\
N_B(\omega_1\omega_2\omega_3)&=&\left[1+n_B(\omega_1)\right]n_B(\omega_2)\left[1 +n_B(\omega_3)\right]
-n_B(\omega_1)\left[1+n_B(\omega_2)\right]n_B(\omega_3).\nonumber
\end{eqnarray}
Here $n_B(\omega)=\left(e^{\omega/T}-1\right)^{-1}$, $n_F(\omega)=\left(e^{\omega/T}+1\right)^{-1}$ with the temperature $T$, the Fourier transforms of the hopping and exchange constants are
\begin{equation}\label{constants}
t_{\bf  k}=4t\gamma_{\bf  k}+4t'\gamma'_{\bf  k},\quad J_{\bf  k}=4J\gamma_{\bf  k},
\end{equation}
$t'$ is the next nearest neighbor hopping constant and $\gamma'_{\bf  k}=\cos(k_x)\cos(k_y)$,
\begin{eqnarray}
B({\bf  k}\omega)&=&-\pi^{-1}{\rm Im}D({\bf  k}\omega),\nonumber\\
A({\bf  k}\omega)&=&-\pi^{-1} \langle\langle a_{\bf  k\sigma}|a^\dagger_{\bf  k\sigma}\rangle\rangle,\label{sfunctions}\\
A'({\bf  k}\omega)&=&-\pi^{-1}\langle\langle a_{\bf  k\uparrow}|a_{\bf  -k\downarrow}\rangle\rangle\nonumber
\end{eqnarray}
are the spectral functions of spin excitations and holes, $\langle\langle a_{\bf  k\sigma}|a^\dagger_{\bf  k\sigma}\rangle\rangle$ and $\langle\langle a_{\bf  k\uparrow}|a_{\bf  -k\downarrow}\rangle\rangle$ are the retarded anticommutator normal and anomalous Green's functions.

In the derivation of Eq.~(\ref{Pi}) we have taken into account that due to the symmetry of Hamiltonian (\ref{Hamiltonian})
$$\langle\langle s^{+1}_{\bf  k}|s^{-1}_{\bf  k}\rangle\rangle=2D({\bf  k}\omega).$$
Notice that the polarization operator (\ref{Pi}) satisfies the following relations:
$${\rm Re}\Pi({\bf  k}\omega)=-{\rm Re}\Pi({\bf  k},-\omega),\quad {\rm Im}\Pi({\bf  k}\omega)={\rm Im}\Pi({\bf  k},-\omega)$$
and hence
$$B({\bf  k}\omega)=-B({\bf  k},-\omega).$$

For $\omega>0$ and zero temperature the first term in the right-hand side of Eq.~(\ref{Pi}) describes the contribution of the process, in which a spin excitation decays to yield a fermion pair and another spin excitation, the second term corresponds to the decay with the formation of three spin excitations, and the third term is connected with the transformation into a fermion pair directly and with assistance of a hole [see Eqs.~(\ref{s2}) and (\ref{ImPi})].

In the below calculations, the following approximate expression for the spectral function of spin excitations is used in the integrands of Eq.~(\ref{Pi}):
\begin{equation}\label{B}
B({\bf  k}\omega)=\frac{2J|C_1|(1-\gamma_{\bf  k})}{\omega_{\bf  k}}\left[\delta(\omega -\omega_{\bf  k})-\delta(\omega+\omega_{\bf  k})\right].
\end{equation}
This expression is derived from Eqs.~(\ref{GfD}) and (\ref{a1a1}) if the damping of spin excitations is neglected to make tractable multiple integrations in Eq.~(\ref{Pi}). In Eq.~(\ref{B}), the term $\omega{\rm Re}\Pi({\bf  k}\omega)$ in the denominator of Eq.~(\ref{GfD}) is also neglected, and the frequency of spin excitations is approximated by Eq.~(\ref{wk}), in which terms containing parameters $F_i$ are dropped for the considered small $x$. The presence of holes exerts main influence on the parameter $\delta$.\cite{Sherman03} In the case of the short-range antiferromagnetic order, which is settled under the influence of thermal fluctuations\cite{Shimahara} and/or a finite hole concentration, the spin-excitation dispersion is gapped at ${\bf  Q}$ (see Fig.~\ref{Fig5} below). The parameter $\delta$ defines the value of this gap, which is directly connected with the spin correlation length. Values of the parameters in Eq.~(\ref{B}) were taken from the self-consistent calculations of Ref.~\onlinecite{Sherman03}.

In the superconducting state, the hole spectral functions for wave vectors near the Fermi surface are approximated by
\begin{eqnarray}
A({\bf  k}\omega)&=&\frac{Z(\xi_{\bf  k}+\varepsilon_{\bf  k})}{2\xi_{\bf  k}}\delta(\omega -\xi_{\bf  k})+\frac{Z(\xi_{\bf  k}-\varepsilon_{\bf  k})}{2\xi_{\bf  k}}\delta(\omega +\xi_{\bf  k})
+\frac{Z'}{8t-2\Delta}S(\omega),\nonumber\\[-1ex]
&&\label{A} \\[-1ex]
A'({\bf  k}\omega)&=&\frac{Z\Delta_{\bf  k}}{2\xi_{\bf  k}}\left[\delta(\omega-\xi_{\bf  k})-\delta(\omega+\xi_{\bf  k})\right],\nonumber
\end{eqnarray}
where
\begin{equation}\label{continuum}
S(\omega)=\left\{\begin{array}{ll}
            1 & \mbox{if $\varepsilon_{\bf  Q}-2t<\omega<-\Delta$,}\\
              & \mbox{or $\Delta<\omega<\varepsilon_{\bf  Q}+6t$,}\\
            0 & \mbox{in other cases,}
\end{array}\right.
\end{equation}
$\varepsilon_{\bf  k}$ and $\xi_{\bf  k}=\sqrt{\varepsilon_{\bf  k}^2+(\Delta_{\bf  k})^2}$ are the hole dispersions in the normal and superconducting states, respectively, with the $d$-wave superconducting gap function $\Delta_{\bf  k}=\Delta[\cos(k_x) -\cos(k_y)]/2$.

The function $S(\omega)$ in $A({\bf  k}\omega)$, Eqs.~(\ref{A}) and (\ref{continuum}), describes the strong HSE continuum, which is typical for the hole spectral function in the $t$-$J$ model. The continuum spans the frequency range approximately equal to the width of the initial band, in the present case $8t$. For moderate doping the lower edge of the continuum lies approximately $2t$ below the bottom of the hole band $\varepsilon_{\bf  k}$,\cite{Sherman99} which is located at the momentum ${\bf  Q}$ in the used hole picture. In accord with the definition of the spectral function the spectral weights of its coherent and incoherent parts are connected by the relation
$$Z+Z'=\frac{1+x}{2}.$$
Below we set $J=0.2t$. This ratio of the parameters belongs to the range derived for hole-doped cuprates.\cite{McMahan,Gavrichkov} For these parameters and hole concentrations $0.07\lesssim x\lesssim 0.15$ the ratio between the spectral weights is $Z/Z'\approx 0.1$.\cite{Sherman99}

We have approximated $S(\omega)$ by constant values within certain frequency ranges to simplify integrations in Eq.~(\ref{Pi}). In Eq.~(\ref{continuum}), the intensity of the HSE continuum is set equal to zero in the interval $-\Delta<\omega<\Delta$. For the considered $d$-wave order parameter this intensity may be nonzero in the ranges $-\Delta<\omega<-|\Delta_{\bf  k}|$ and $|\Delta_{\bf  k}|<\omega<-\Delta$. However, since the hole damping decreases as the Fermi level is approached, it can be supposed that the continuum intensity in these ranges is much smaller than far apart from $\omega=0$. Notice, however, that its perceptible intensity in these intervals may be of critical importance for the character of the magnetic response below the resonance frequency. In the calculations we increased the continuum intensity in these regions up to the value it has far from the Fermi level. The response became commensurate due to the growth of $|{\rm Im}D|$ at the antiferromagnetic wave vector.

For the considered hole dispersions corresponding to moderate $x$ and for frequencies $\omega\lesssim 0.3t=150$~meV the main contribution to the magnetic susceptibility from coherent peaks of the spectral functions (\ref{A}) is made by the last term in the right-hand side of Eq.~(\ref{Pi}) (here and below in going to energy units we proceed from the estimate\cite{McMahan,Gavrichkov} $t=500$~meV). As mentioned above, this term is connected with the process of the spin-excitation decay into fermion pairs. The substitution of Eq.~(\ref{A}) into (\ref{Pi}) gives an addend with the multiplier $\delta(\omega-\xi_{\bf  k+q}-\xi_{\bf  q})$ in the imaginary part of this term. The approximation used for the representation of this $\delta$-function has an important impact on the character of the magnetic response below the resonance frequency. In fact, this approximation is connected with the shape of the coherent peaks in the hole spectral function. If this $\delta$-function is approximated by a Lorentzian with a constant width $\Gamma$, the low-frequency incommensurate response arises only for a Fermi surface with nesting, as it was noted for the analogous term in the itinerant electron theory.\cite{Lavagna,Bulut,Norman} However, not all dispersions derived from the photoemission of $p$-type cuprates possess nesting. Nevertheless, the incommensurate magnetic response is observed also in crystals with such dispersions. As indicated above, the problem is connected with far tails of the Lorentzian. They strengthen ${\rm Im}\chi$ at the antiferromagnetic wave vector. This hides incommensurate peaks, which are less intensive in the absence of nesting. However, the decay of tails of the spectral peaks has to proceed faster than in Lorentzian, since the damping of the respective states grows with distance from the Fermi level. Besides, the tails have to be cut off in the superconducting gap. To take these facts into account, we approximate the function $\delta(\omega-\xi_{\bf  k+q}-\xi_{\bf  q})$ by the $\Pi$-shaped step function
\begin{equation}\label{stepfunction}
P(\omega,\Gamma)=\left\{\begin{array}{cl}
            (2\Gamma)^{-1} & \mbox{if $-\Gamma<\omega<\Gamma$,}\\
            0 & \mbox{in other cases.}
\end{array}\right.
\end{equation}
With this approximation, the low-frequency incommensurate response was obtained for several hole dispersions used earlier for $p$-type cuprates,\cite{Brinckmann,Norman,Radtke,Schabel,Macridin} including those derived from photoemission data.

For $\omega>0$ and $T=0$ the result of the substitution of spectral functions (\ref{B}) and (\ref{A}) into Eq.~(\ref{Pi}) is given in Eq.~(\ref{ImPi}), which was used in the present calculations.

\section{Results and discussion}
\begin{figure}[t]
\centerline{\includegraphics[width=5cm]{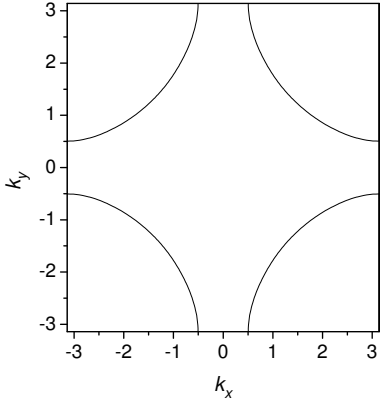}}
\caption{The Fermi surface of the dispersion (\protect\ref{dispersion1}).} \label{Fig1}
\end{figure}

In this section, figures were obtained with the following dispersions for the coherent part of the hole spectrum:
\begin{equation}
\frac{\varepsilon_{\bf  k}}{t}=\frac{1}{2}\Big[\cos(k_x)+\cos(k_y)\Big]
-0.3\times\cos(k_x)\cos(k_y)-0.2,\label{dispersion1}
\end{equation}
\begin{eqnarray}
\varepsilon_{\bf  k}&=&-0.1305+0.5951\times\frac{1}{2}\Big[\cos(k_x)+\cos(k_y)\Big]
-0.1636\times\cos(k_x)\cos(k_y)\nonumber\\
&&+0.0519\times\frac{1}{2}\Big[\cos(2k_x)+\cos(2k_y)\Big]     +0.1117\times\frac{1}{2}\Big[\cos(2k_x)\cos(k_y)\nonumber\\
&&+\cos(k_x)\cos(2k_y)\Big]
-0.0510\times\cos(2k_x)\cos(2k_y).\label{dispersion2}
\end{eqnarray}
Dispersions similar to that given by Eq.~(\ref{dispersion1}) are frequently used for the description of hole bands in $p$-type cuprates (see, e.g., Refs.~\onlinecite{Brinckmann,Radtke,Macridin}). Its Fermi surface is shown in Fig.~\ref{Fig1}.
The second dispersion (\ref{dispersion2}) was borrowed from Ref.~\onlinecite{Norman}. It was derived from the photoemission data in Bi$_2$Sr$_2$CaCu$_2$O$_8$. Notice that if band (\ref{dispersion1}) and similar bands in Refs.~\onlinecite{Brinckmann,Radtke,Macridin} were used for underdoped cuprates, $x\lesssim 0.12$, the dispersion (\ref{dispersion2}) corresponds to $x\approx 0.17$.\cite{Norman} The coefficients in Eq.~(\ref{dispersion2}) are in electronvolts. In this equation, signs are opposite to those in Ref.~\onlinecite{Norman}, since we use the hole picture. The general shape of the Fermi surface for dispersion (\ref{dispersion2}) is similar to that shown in Fig.~\ref{Fig1}.

\begin{figure}[t]
\centerline{\includegraphics[width=5cm]{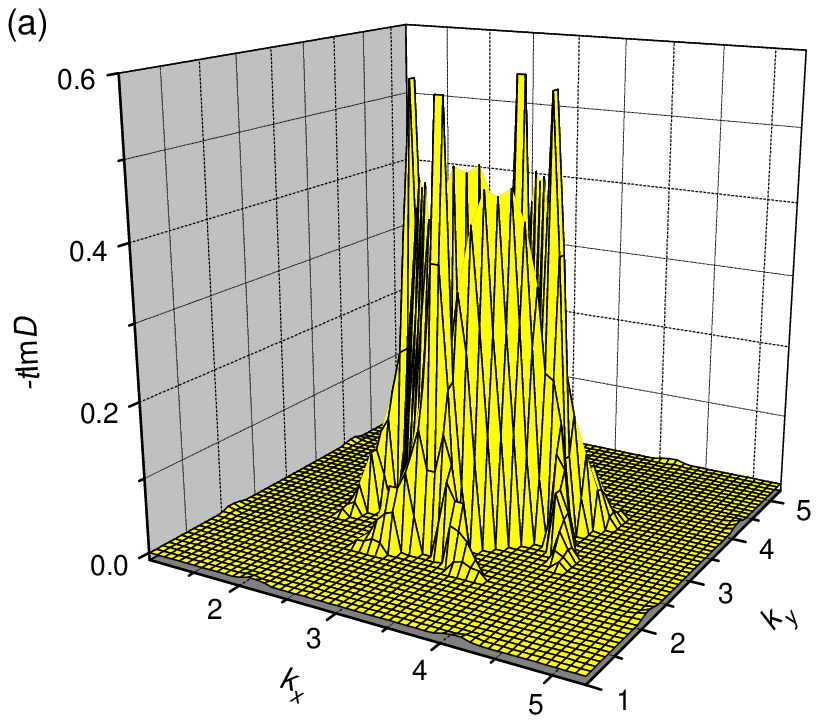}}\vspace{2ex}
\centerline{\includegraphics[width=5cm]{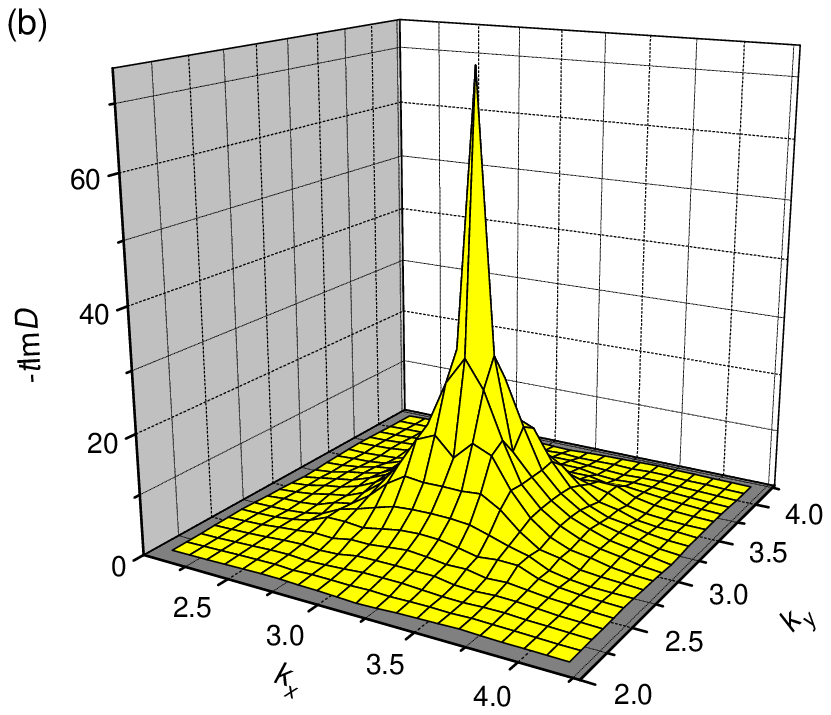}}\vspace{2ex}
\centerline{\includegraphics[width=5cm]{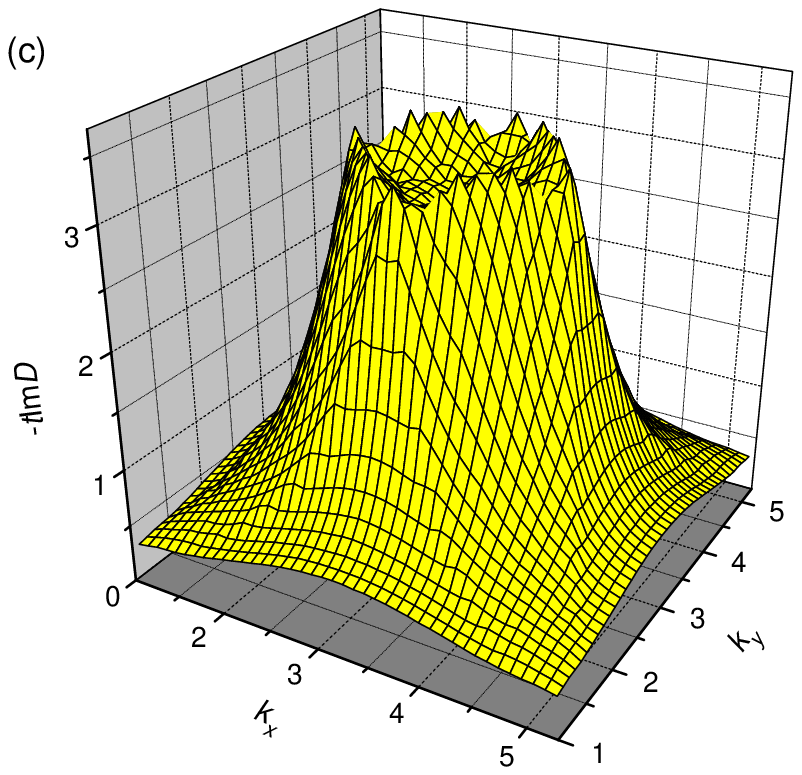}}
\caption{The imaginary part of the spin Green's function for the hole dispersion~(\protect\ref{dispersion1}); $\omega=0.04t=20$~meV (a), $0.077t=38.5$~meV (b), and $0.12t=60$~meV (c). Other parameters are given in the text.} \label{Fig2}
\end{figure}

\begin{figure}
\centerline{\includegraphics[width=7cm]{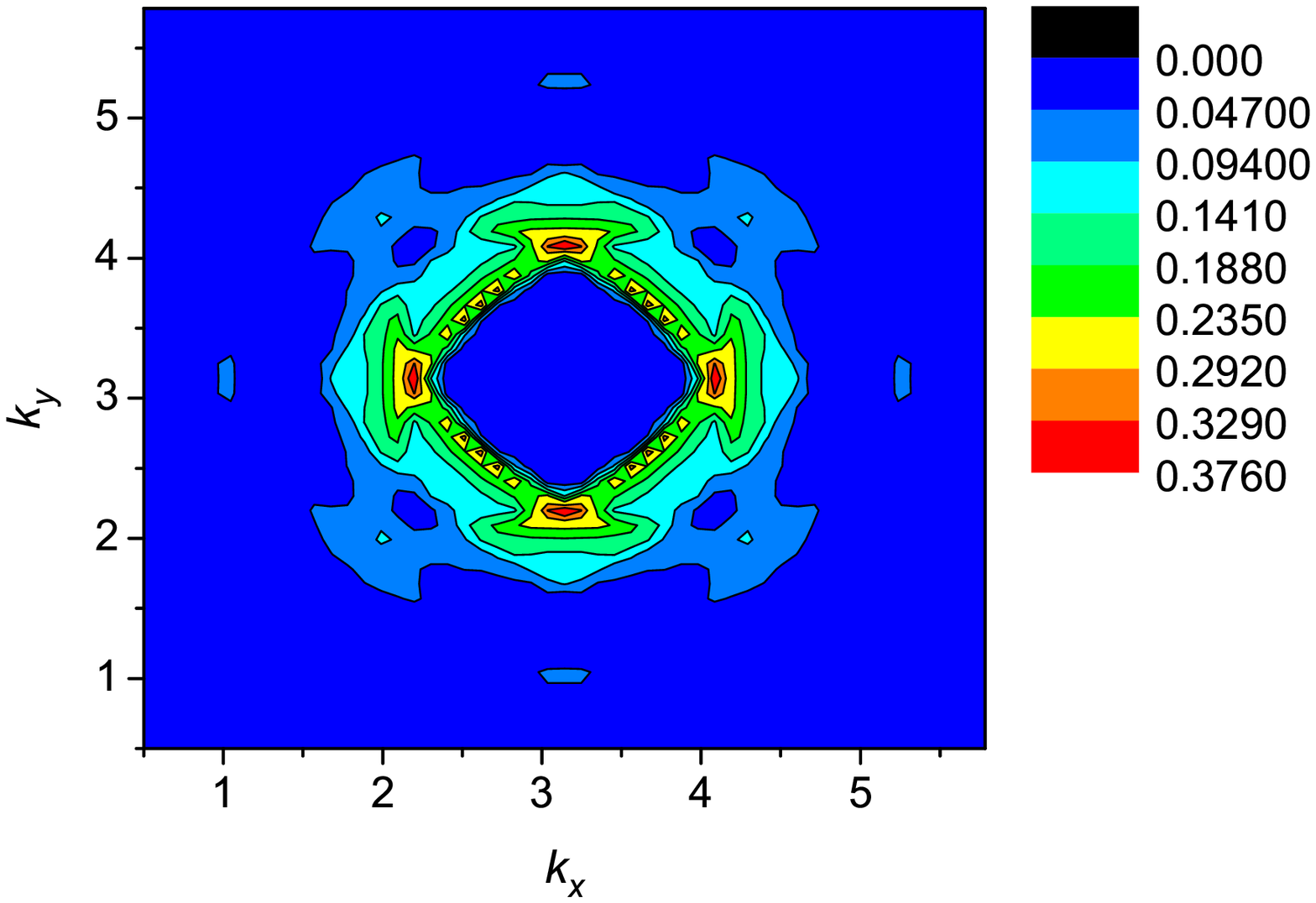}}\vspace{3ex}
\hspace*{-1.7ex}\centerline{\includegraphics[width=6.75cm]{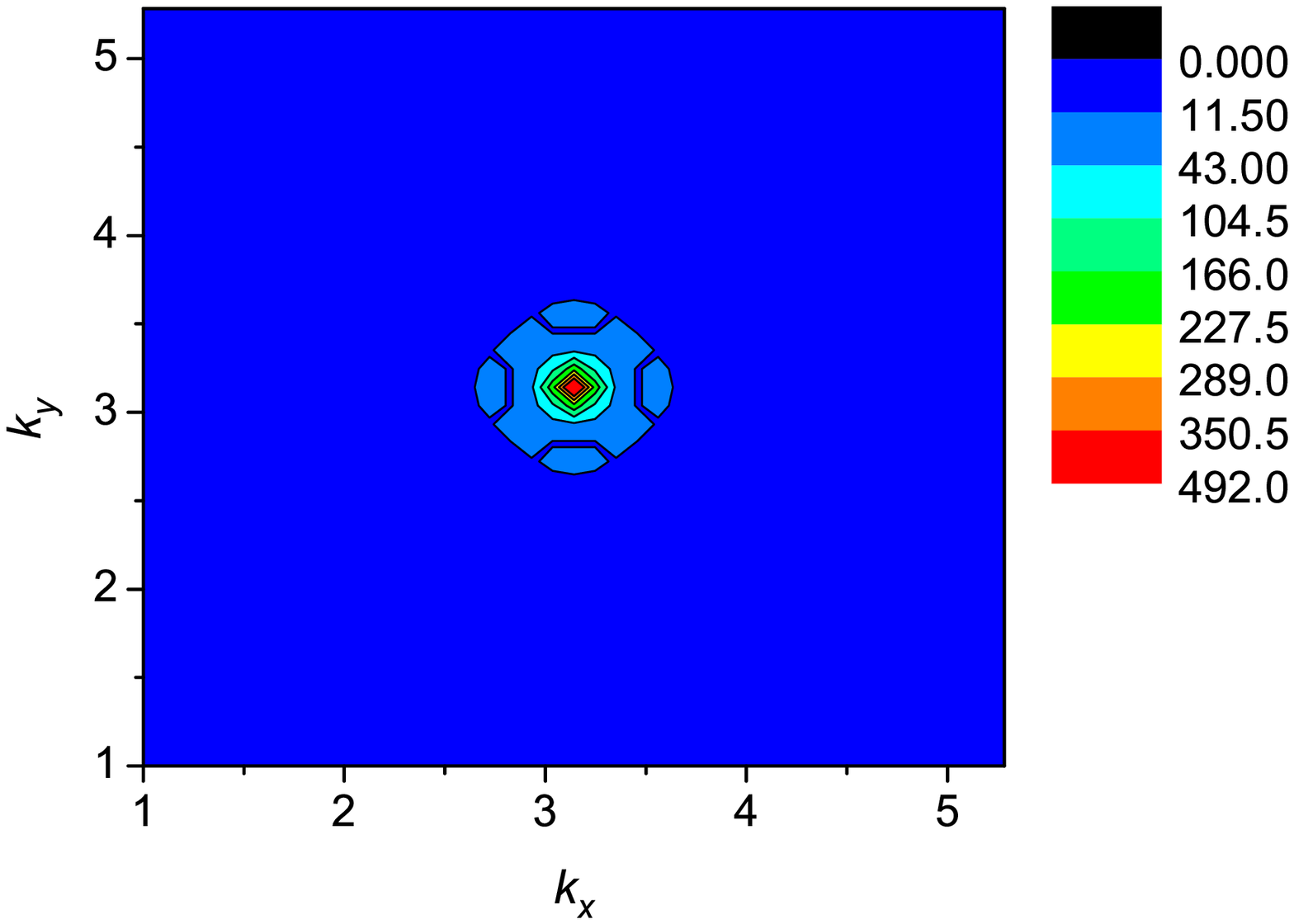}}\vspace{3ex}
\hspace*{-0.8ex}\centerline{\includegraphics[width=6.9cm]{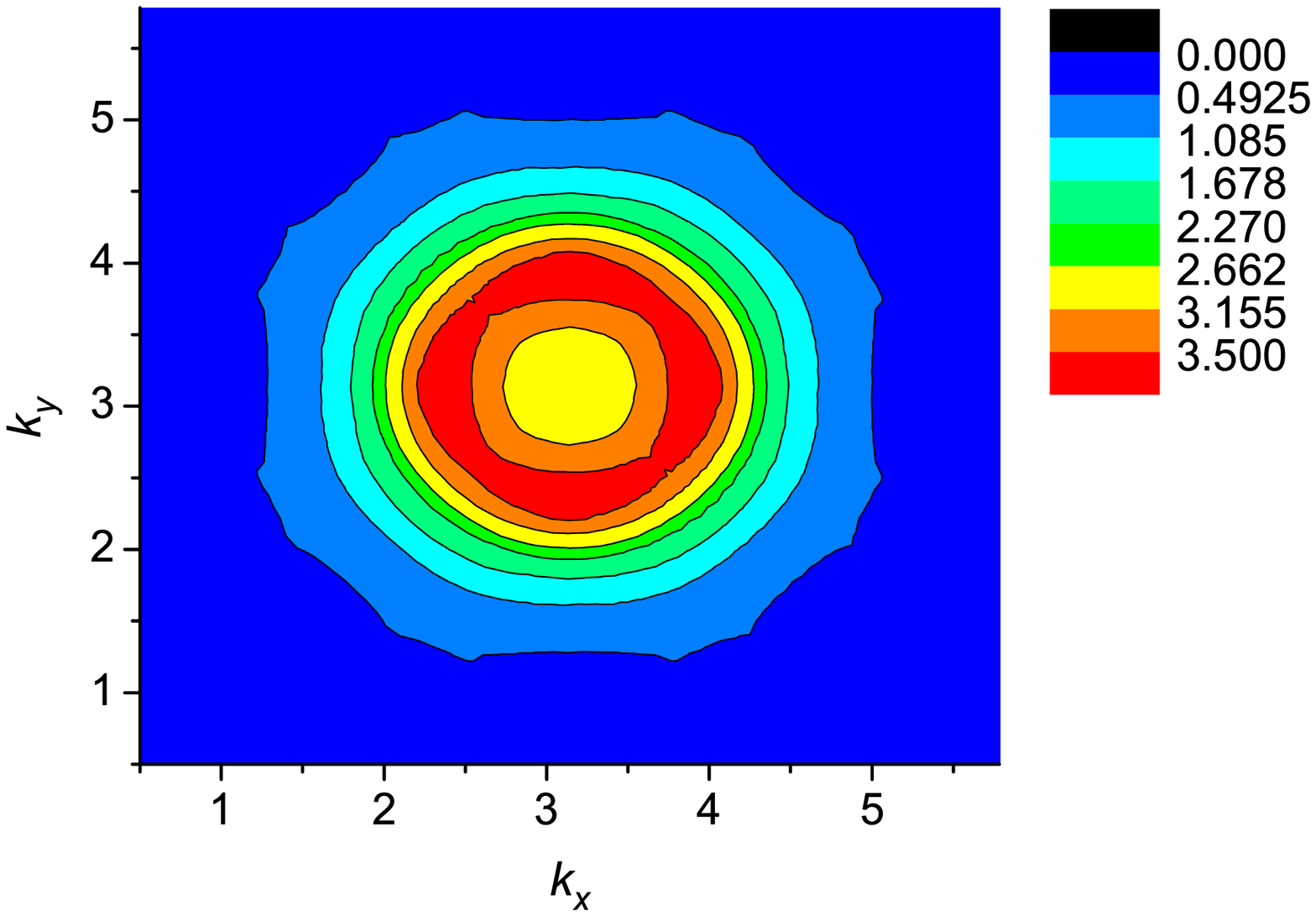}}
\caption{The contour plots of $-t{\rm Im}D({\bf  k}\omega)$ for the hole dispersion~(\protect\ref{dispersion2}); $\omega=0.06t=30$~meV (a), $0.0747t=37.4$~meV (b), and $0.1t=50$~meV (c). Other parameters are given in the text.} \label{Fig3}
\end{figure}

The other parameters used in the below calculations are the following: $J/t=0.2$, $Z/Z'=0.1$, $\Gamma=0.002t$, $\delta=0.25$, $\Delta=0.05t$, $C_1=-0.2$, and $\alpha=1.7$. The last two parameters are taken from the self-consistent calculations for low hole concentrations\cite{Sherman03} and are close to the results for the undoped case.\cite{Shimahara} The parameter $\delta$ is somewhat increased in comparison with the self-consistent result of Ref.~\onlinecite{Sherman03} to compensate the correction introduced by the term $\omega{\rm Re}\Pi({\bf  k}\omega)$ in the present calculations. We chose for $\delta$ the value, which provides the spin gap $\omega'_{\bf  Q}=\left[\omega^2_{\bf  Q}+\omega'_{\bf  Q}{\rm Re}\Pi({\bf  Q}\omega'_{\bf  Q})\right]^{1/2}\approx 40$~meV with dispersion (\ref{dispersion1}). Gaps of such a magnitude were obtained in the self-consistent calculations for moderate hole concentrations.\cite{Sherman03} This value is also close to the frequency of the resonance peak observed in some cuprates.\cite{Bourges98,Rossat} Results of calculations with Eqs.~(\ref{GfD}) and (\ref{ImPi}) are given in Fig.~\ref{Fig2} for dispersion (\ref{dispersion1}) and in Fig.~\ref{Fig3} for dispersion (\ref{dispersion2}).

As seen in the figures, for both dispersions the magnetic response is incommensurate for small frequencies (Figs.~\ref{Fig2}(a) and \ref{Fig3}(a)), and the maxima are located on axes of the Brillouin zone. With increasing frequency the response becomes commensurate (Figs.~\ref{Fig2}(b) and \ref{Fig3}(b)) and reaches the maximal intensity. For even larger frequencies, the response is again incommensurate (Figs.~\ref{Fig2}(c) and \ref{Fig3}(c)). However, this time it looks more isotropic than for small frequencies. Similar behavior of the magnetic susceptibility are observed experimentally in $p$-type cuprates (see, e.g., Ref.~\onlinecite{Tranquada}). For frequencies $\omega>0.26t=130$~meV for dispersion (\ref{dispersion1}) and $\omega>0.2t=100$~meV for dispersion (\ref{dispersion2}) with the used set of parameters the momentum dependence of the susceptibility transforms to a broad commensurate maximum. For both dispersions for frequencies $\omega\lesssim 0.02t=10$~meV the susceptibility is peaked at the diagonals of the Brillouin zone.

Notice that in Ref.~\onlinecite{Norman} no incommensurability was found for small frequencies with dispersion (\ref{dispersion2}), in contrast to the result shown in Fig.~\ref{Fig3}(a). As mentioned above, this discrepancy is connected with different representations of the $\delta$-functions ensuring the energy conservation in the convolutions of coherent parts of the hole spectral functions. With representation (\ref{stepfunction}) results similar to those shown in Figs.~\ref{Fig2} and \ref{Fig3} were also obtained for some other bands used earlier  for the description of $p$-type cuprates. Susceptibilities obtained with different dispersions vary somewhat in the location and intensity of maxima and in some other details. Such discrepancies can be expected, since the dispersions were obtained for crystals of different compositions and with dissimilar hole concentrations. For the bonding band from Ref.~\onlinecite{Schabel} we found low-frequency maxima on diagonals of the Brillouin zone, while for other considered bands they are on the axes, at least for not too small frequencies.

\begin{figure}[t]
\centerline{\includegraphics[width=6cm]{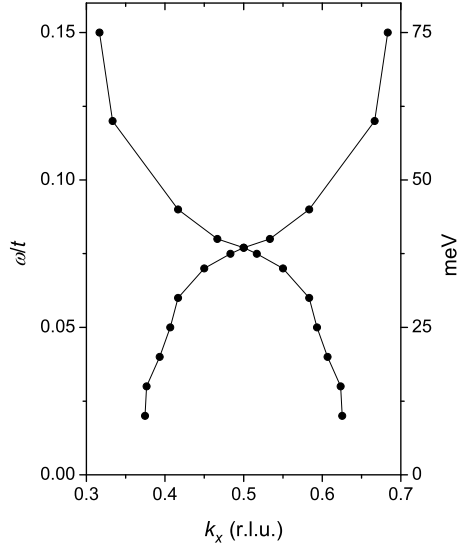}}
\caption{The dispersion of the susceptibility maxima along the axis of the Brillouin zone for the parameters of Fig.~\protect\ref{Fig2}. The momentum is in the reciprocal lattice units.} \label{Fig4}
\end{figure}

For band (\ref{dispersion1}) the dispersion of susceptibility maxima, which are located on the $X$ axis of the Brillouin zone, is shown in Fig.~\ref{Fig4}. It is similar to the hourglass dispersion observed experimentally.\cite{Tranquada} Notice that in Fig.~\ref{Fig4} the wave vector is in the reciprocal lattice units, $2\pi/a=1$~r.l.u. As seen in the figure, for dispersion (\ref{dispersion1}) the maxima are located at $k_x=\frac{1}{2}\pm \frac{1}{8}$ at low frequencies, which is typical for the hole concentration $x=\frac{1}{8}$.\cite{Tranquada}

\begin{figure}[t]
\centerline{\includegraphics[width=7cm]{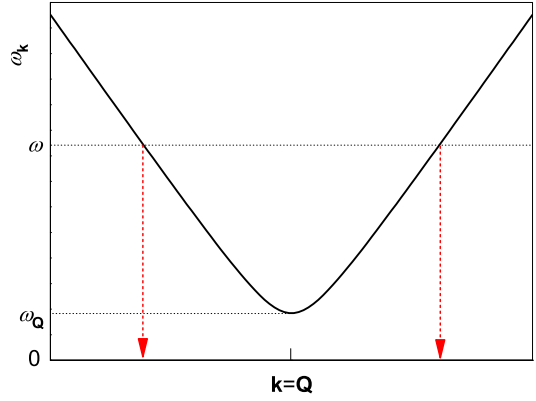}}
\caption{The dispersion of spin excitations (\protect\ref{wk}) near the antiferromagnetic wave vector ${\bf  Q}$. Wave vectors, for which the resonance condition (\protect\ref{resonance}) is satisfied, are shown by arrows.} \label{Fig5}
\end{figure}

The formation mechanisms of the incommensurability below and above the resonance frequency are different. Green's function (\ref{GfD}) has the resonance denominator, which is responsible for the incommensurate response above $\omega_r$. If we drop temporarily the real part of the polarization operator in the denominator of Eq.~(\ref{GfD}) the resonance condition reads
\begin{equation}\label{resonance}
\omega=\omega_{\bf  k}.
\end{equation}
As follows from Eq.~(\ref{wk}), near ${\bf  Q}$ the spin-excitation dispersion is given by the expression
\begin{equation}\label{wQ}
\omega_{\bf  k}\approx\sqrt{\omega_{\bf  Q}^2+c^2({\bf  Q}-{\bf  k})^2},\quad c=\sqrt{8\alpha|C_1|}J.
\end{equation}
This dispersion is shown in Fig.~\ref{Fig5}. As mentioned above, the dispersion has a gap at ${\bf  Q}$ and the magnitude of the gap $\omega_{\bf  Q}$ is connected with the correlation length of the short-range antiferromagnetic order.\cite{Shimahara,Sherman03} For $\omega>\omega_{\bf  Q}$ the graphic solution of Eq.~(\ref{resonance}) is shown in Fig.~\ref{Fig5}. Thus, if the imaginary part of the polarization operator is not too large, the susceptibility is peaked at momenta indicated by arrows in the figure, i.e. on the circle surrounding ${\bf  Q}$,\cite{Barzykin,Sherman03a} as seen in Figs.~\ref{Fig2}(c) and \ref{Fig3}(c). This is qualitatively the mechanism of the formation of the incommensurate response above the hourglass neck. The inclusion of ${\rm Re}\Pi({\bf  k}\omega)$ leads to a renormalization of $\omega_{\bf  Q}$ and dispersion (\ref{wQ}) immediately above the spin gap. For larger frequencies the real part of the denominator in Eq.~(\ref{GfD}) becomes nonzero. Besides, the damping ${\rm Im}\Pi({\bf  k}\omega)$ grows rapidly with $\omega$. As a consequence the high-frequency incommensurate maxima are less pronounced or a broad commensurate peak is formed. Such shapes of the susceptibility were also observed experimentally.\cite{Stock}

\begin{figure}[t]
\centerline{\includegraphics[width=7cm]{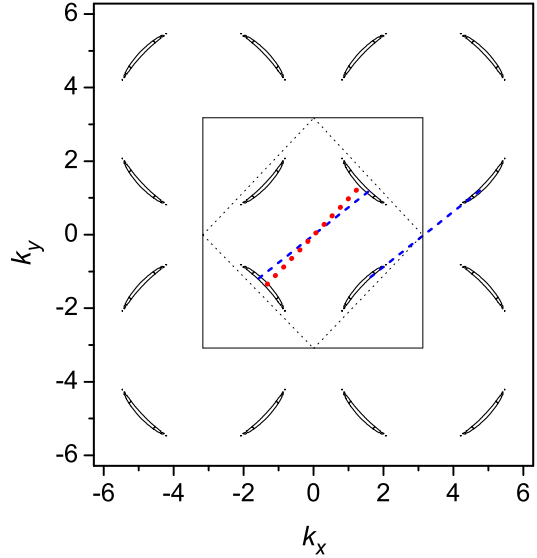}}
\caption{The contour plot of the dispersion $\xi_{\bf  k}=\sqrt{\varepsilon_{\bf  k}^2+\Delta_{\bf  k}^2}$, where $\varepsilon_{\bf  k}$ is given by Eq.~(\protect\ref{dispersion1}) and $\Delta=0.05t$. Contours correspond to the energy $0.03t$. The squares shown by thin solid and dotted lines are the first and magnetic Brillouin zones. The blue dashed lines and the red dotted line connect fermion pairs contributing to the susceptibility on the axis and diagonal of the momentum space, respectively.} \label{Fig6}
\end{figure}
Below the resonance frequency the momentum dependence of the imaginary part of the susceptibility is determined by ${\rm Im}\Pi({\bf  k}\omega)$ in the numerator of the expression
\begin{equation}\label{ImD}
{\rm Im}D({\bf  k}\omega)=
\frac{h_{\bf  k}\omega{\rm Im}\Pi({\bf  k}\omega)}{\left[\omega^2-\omega{\rm Re}\Pi({\bf  k}\omega)-\omega_{\bf  k}^2\right]^2+\left[\omega{\rm Im}\Pi({\bf  k}\omega)\right]^2}.
\end{equation}
In this frequency range, the last sum in Eq.~(\ref{Pi}) [which corresponds to the last sum in Eq.~(\ref{ImPi})] makes the main contribution to the polarization operator. This sum is connected with the decay of a magnetic excitation into two fermions. The incommensurate peaks in the susceptibility are related to the contribution of coherent parts of the hole spectral functions into this sum [the term with the multiplier $P(\omega-\xi_{\bf  k+q}-\xi_{\bf  q})$ in the last sum of Eq.~(\ref{ImPi})]. The fermions with energies $\xi_{\bf  k}\approx\omega/2$ play a key part here. The location of these states in the momentum space is shown in Fig.~\ref{Fig6}. In this figure, crescent pockets are sections of the fermion dispersion $\xi_{\bf  k}=\sqrt{\varepsilon_{\bf  k}^2+\Delta_{\bf  k}^2}$ at the energy $0.03t$. For small energies this dispersion looks like flattened cones originating at the node points. The blue dashed and red dotted lines connect pairs of fermion states, which give the main contribution to the magnetic susceptibility on an axis and diagonal of the momentum space, respectively. The location of these states is restricted to the crescent pockets, and the difference in their wave vectors is equal to the spin excitation momentum. As a consequence the low-frequency susceptibility is sharply peaked at certain regions of the Brillouin zone, as seen in Figs.~\ref{Fig2}(a) and \ref{Fig3}(a). The line segments shown by dashed and dotted lines in Fig.~\ref{Fig6} differ from the antiferromagnetic wave vector and, therefore, the response is incommensurate (in this figure, opposite sides of the magnetic Brillouin zone are displaced by the vector ${\bf  Q}$ from each other). This is the mechanism of the formation of the incommensurate response below the resonance frequency. Notice that it implies the fulfillment of the energy conservation: $\omega=\xi_{\bf  q}+\xi_{\bf  k+q}$ [see the mentioned term in Eq.~(\ref{ImPi})]. In the processes, in which distant tails of the coherent peaks contribute to the polarization operator, this energy conservation is violated. If this contribution dominates the low-frequency response will be commensurate.

For frequencies immediately below $\omega_r$ a number of fermion pairs, which contribute to the susceptibility peak on an axis of the Brillouin zone, is larger than for the maximum on a diagonal. This conclusion can be drawn from Fig.~\ref{Fig6} --  there are two groups of fermion pairs, which make the main contribution in the former case, and only one group in the latter. This explains the maxima of $-{\rm Im}D({\bf  k}\omega)$ on the axes in this frequencies range. The same location of susceptibility maxima is observed in experiment.\cite{Yoshizawa,Birgeneau,Dai,Arai,Bourges} However, with decreasing $\omega$ the size of the crescent pockets is reduced and the intensity on axes goes down faster than on diagonals in our calculations. Finally, at very small frequencies the susceptibility becomes peaked at the diagonals of the Brillouin zone (for parameters of Figs.~\ref{Fig2} and \ref{Fig3} it happens at $\omega=0.02t$). For even smaller frequencies the intensity on the axes disappears at all. The analogous behavior was observed in the itinerant-electron approach.\cite{Brinckmann} This result seems contrary to experimental data, where the susceptibility is peaked on the axes up to $\omega=0$.\cite{Katano,Lake,Kofu,Haug} A possible way to resolve this contradiction is discussed in the next section.

With increasing the hole concentration regions of the Brillouin zone, which are filled by holes, grow.\cite{Ino} These regions are revealed by the crescent pockets around the zone corners in Fig.~\ref{Fig6}. As a consequence dashed and dotted line segments in the figure move more and more away from ${\bf  Q}$. Thus, the incommensurability parameter $\kappa$ grows with $x$. This growth is also promoted by the concentration dependence of the spin-gap frequency $\omega_{\bf  Q}$. As follows from Eq.~(\ref{ImD}), for small frequencies
$${\rm Im D}({\bf  k}\omega)\approx \omega h_{\bf  k}\omega_{\bf  k}^{-4}{\rm Im}\Pi({\bf  k}\omega).$$
In this equation, the multiplier $\omega_{\bf  k}^{-4}$ is peaked at ${\bf  Q}$ [see Fig.~\ref{Fig5} and Eq.~(\ref{wQ})] and therefore it impedes the appearance of the incommensurate peaks. However, the gap $\omega_{\bf  Q}$ grows with $x$,\cite{Sherman03} effectively slackening the momentum dependence of $\omega_{\bf  k}^{-4}$ and promoting the separation of the susceptibility maxima from the antiferromagnetic wave vector. The indicated increase of the incommensurability parameter with $x$ is observed experimentally.\cite{Yamada}

\begin{figure}[t]
\centerline{\includegraphics[width=7cm]{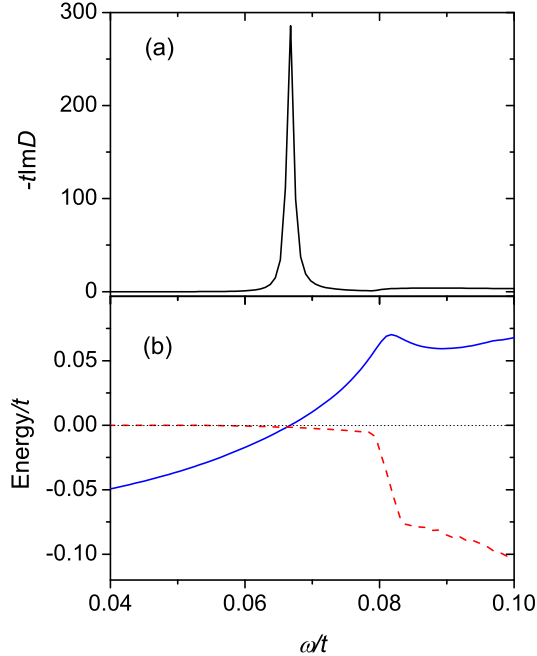}}
\caption{(a) The resonance peak in the spin susceptibility. (b) The real (the blue solid line) and imaginary (the red dashed line) parts of the denominator of Eq.~(\protect\ref{GfD}). Parameters are the same as in Fig.~\protect\ref{Fig2}, except that $\delta=0.16$.}\label{Fig7}
\end{figure}
As seen in Figs.~\ref{Fig2} and \ref{Fig3}, for the used parameters in the region of the hourglass neck the susceptibility peaks are much more intensive and narrower than the maxima for smaller and larger frequencies. This peculiarity of the spin-excitation spectrum is called the resonance peak.\cite{Rossat} The formation mechanism of the peak is explained in Fig.~\ref{Fig7}(b), in which the real and imaginary parts of the denominator of the spin Green's function (\ref{GfD}) are shown. The considered crystals have two energy parameters, which may have comparable values. The first of these parameters is the spin-gap frequency $\omega'_{\bf  Q}$, for which the real part of the mentioned denominator vanishes at ${\bf  k=Q}$. The second parameter is the lower edge of the two-fermion continuum, $\omega_e({\bf  Q})=\min(\xi_{\bf  q}+\xi_{\bf  Q+q})$ at the antiferromagnetic momentum. Due to a small but finite damping of fermion states this edge appears as a fuzzy step near $0.08t$ in the imaginary part of the denominator in Fig.~\ref{Fig7}(b). This step has the same origin as in the theory of itinerant carriers,\cite{Brinckmann,Eremin05,Abanov} where for a vanishingly small damping it transforms to a discontinuous jump leading to a logarithmic singularity in the real part of the noninteracting susceptibility. In this theory, the resonance peak is related to this singularity. As seen in Fig.~\ref{Fig7}(b), the small damping $\Gamma=0.002t=1$~meV transforms the singularity into the weak maximum. For the parameters of Fig.~\ref{Fig7}, as well as for Figs.~\ref{Fig2} and \ref{Fig3}, $\omega'_{\bf  Q}<\omega_e({\bf  Q})$, the real part of the denominator of Eq.~(\ref{GfD}) vanishes in the region of small damping below $\omega_e({\bf  Q})$, which produces the narrow and intensive resonance peak.\cite{Sherman03a} Apparently in this case $\omega_r$ coincides with the spin-gap frequency $\omega'_{\bf  Q}$. It grows with $x$, as observed experimentally in YBa$_{2}$Cu$_{3}$O$_{7-y}$,\cite{Bourges98} since with increasing the hole concentration the spin correlation length decreases. Notice that the weak damping below $\omega_e({\bf  Q})$ is mainly connected with the contribution of the HSE continuum [the second and third terms in the last sum of Eq.~(\ref{ImPi})]. As the wave vector recedes from ${\bf  Q}$ the spin-excitation frequency $\omega'_{\bf  k}$ crosses the edge of the electron-hole continuum $\omega_e({\bf  k})$, which separates regions of weak and strong damping. The incommensurate peaks become broader and less intensive and finally the magnetic response becomes commensurate, as mentioned above. The considered case resembles the situation in YBa$_2$Cu$_3$O$_{7-y}$ and Bi$_2$Sr$_2$CaCu$_2$O$_{8+\delta}$.\cite{Dai,Arai,Bourges,Bourges98}

Another situation occurs if $\omega'_{\bf  Q}>\omega_e({\bf  Q})$. Due to a larger damping the susceptibility maximum at ${\bf  Q}$ is broader and less intensive than in the previous case. In particular, this situation takes place for $\Delta=0.03t$ if the other parameters are the same as for Fig.~\ref{Fig2}. Due to broader maxima the hourglass dispersion looks somewhat different than that shown in Fig.~\ref{Fig4} -- the neck with commensurate susceptibility maxima occupies a substantial range of frequencies (for the mentioned parameters from $0.05t$ to $0.12t$). The broad maximum with a moderate peak intensity at the antiferromagnetic wave vector resembles the picture observed in lanthanum cuprates.\cite{Haiden,Tranquada}

\section{Conclusion}
The above discussion was referred to the superconducting state. To consider the pseudogap phase, we have to drop the contribution of the anomalous hole Green's functions and to introduce a finite damping in the pseudogap region $|\omega|<\Delta$. As mentioned above, the damping blurs the incommensurate peaks or even leads to a commensurate response. Such changes in the susceptibility are observed experimentally above $T_c$.\cite{Kofu} Additionally it should be taken into account that in the pseudogap phase, the depression of the hole spectral intensity near the Fermi level is not described by the $d$-wave function. In this state, near the diagonals of the Brillouin zone gapless regions of the Fermi surface -- arcs -- are located.\cite{Shen,Kanigel} As mentioned above, in the superconducting phase with decreasing frequency the intensity of the susceptibility maxima on the axes goes down more rapidly than on the diagonals. As a result for small frequencies the intensity is peaked at diagonals. This fact is connected with shrinking crescent pockets of fermion states, which contribute to the susceptibility maxima, when $\omega\rightarrow 0$. Obviously, with the appearance of arcs in the pseudogap phase the situation is changed: with decreasing frequency the pockets do not shrink to points on the diagonals but rather acquire the shape of the arcs.
Consequently, it can be expected that the susceptibility maxima will be located on the axes up to zero frequency. The incommensurate elastic response with maxima peaked on the axes is observed in lanthanum and yttrium cuprates.\cite{Katano,Lake,Kofu,Haug} Based on the sharp difference in the spin correlation lengths below and above the energy $E_g\approx 4$~meV it was supposed\cite{Kofu} that below $T_c$ the crystal La$_{1.875}$Sr$_{0.125}$CuO$_4$ is phase-separated into superconducting and non-superconducting regions. From the above discussion one can expect that the incommensurate elastic response is due to the latter regions, which are in the pseudogap state, and both regions contribute to the inelastic susceptibility with the hourglass dispersion for $\omega>E_g$. This picture allows one to suggest a mechanism for the strengthening of the incommensurate elastic response by magnetic fields $H<H_{c2}$, observed\cite{Katano,Lake,Haug} in lanthanum and yttrium cuprates. The growing field increases the part of the crystal occupied by vertex cores, which are in the pseudogap state and contribute to the response.

In summary, we calculated the magnetic susceptibility of the two-dimensional $t$-$J$ model. To consider the case of strong electron correlations the Mori projection operator technique was used. All processes contributing to the spin polarization operator -- the decay into a fermion pair, the decay into the pair assisted by a hole or with a creation of a spin excitation, and the decay into three spin excitations -- were taken into account. In these calculations, both the coherent and the strong incoherent parts of the hole spectral function were allowed for. For the coherent part several dispersions found in the literature were analyzed, including those designed to fit  photoemission data. For the superconducting state with the $d$-wave order parameter we found that at certain conditions all considered dispersions produce the incommensurate magnetic response below the resonance frequency. For this incommensurability the nesting of the Fermi surface is not required. This result proposes the explanation for the fact that the low-frequency incommensurability is observed in a wide range of hole concentrations and in different families of $p$-type cuprates. The mentioned conditions reduce to the fast decay of the tails of the hole coherent peaks and to a small intensity of the hole-spin-excitation continuum near the Fermi level. Both conditions are related to the damping of hole states, which decreases on approaching the Fermi level. The used equations allowed us to obtain the hourglass dispersion of the susceptibility maxima. Their locations in the momentum space are close to those observed experimentally. The upper part of the hourglass dispersion is related to the dispersion of spin excitations. The part below the resonance frequency is connected with the incommensurate maxima of spin-excitation damping, which is mainly caused by the decay into a fermion pair in this frequency range. Immediately below the resonance frequency the susceptibility is sharply peaked at the axes of the Brillouin zone due to a larger number of final states in comparison with the diagonals. The intensity distribution in the upper part of the hourglass dispersion is more isotropic. In this picture, the growth of the low-frequency incommensurability parameter with doping is related to the increasing area occupied by holes in the Brillouin zone and to the rising gap in the spin-excitation dispersion at the antiferromagnetic momentum. The resonance peak is narrow and intensive if the gap frequency falls on the region of small damping below the edge of the two-fermion continuum at the antiferromagnetic momentum. In this case, the resonance frequency is close to the gap frequency. If the gap frequency exceeds the edge of the two-fermion continuum, a broader and less intensive commensurate maximum is produced.

\appendix*
\section{}
In this Appendix equations of Sec.~II are derived. In the Mori projection operator technique $D({\bf  k}\omega)$ is calculated from the Kubo relaxation function
\begin{equation}\label{Kubo}
\left(\left(s^z_{\bf  k}|s^z_{\bf  -k}\right)\right)=\int^\infty_0 d t e^{i\omega t}\int^\infty_t d t'\left\langle \left[s^z_{\bf  k}(t'),s^z_{\bf  -k}\right]\right\rangle.
\end{equation}
The two functions are connected by the relation
\begin{equation}\label{KuboGreen}
\omega\left(\left(s^z_{\bf  k}|s^z_{\bf  -k}\right)\right)=\left(s^z_{\bf  k},s^z_{\bf  -k}\right)+D({\bf  k}\omega),
\end{equation}
where
\begin{equation}\label{product}
\left(s^z_{\bf  k},s^z_{\bf  -k}\right)=i\int^\infty_0 d t\left\langle\left[s^z_{\bf  k}(t),s^z_{\bf  -k}\right]\right\rangle.
\end{equation}

In this approach, Green's and relaxation functions are represented by continued fractions. The elements of these fractions can be calculated using the recurrence relation,\cite{Sherman87} which resembles that of the Lanczos algorithm.\cite{Cullum} As applied to the relaxation function (\ref{Kubo}), the continued fraction and the recurrence relation read
\begin{equation}\label{cfraction}
\left(\left(s^z_{\bf  k}|s^z_{\bf  -k}\right)\right)=\frac{\left(s^z_{\bf  k},s^z_{\bf  -k}\right)}{\omega-E_0-\frac{\textstyle V_0}{\textstyle\omega-E_1-V_1R({\bf  k}\omega)}},
\end{equation}
\begin{eqnarray}
&&[A_n,H]=V_{n-1} A_{n-1}+E_n A_n+A_{n+1},\;
E_n=\left([A_n,H],A_n^\dagger\right)\left(A_n,A_n^\dagger\right)^{-1},
\nonumber\\[-1ex]
&&\label{Lanczos}\\[-1ex]
&&V_n=\left(A_{n+1},A_{n+1}^\dagger\right)\left(A_n,A_n^\dagger\right)^{-1},
\; V_{-1}=0,\; A_0=s^z_{\bf  k},\; n=0,1,2\ldots,\nonumber
\end{eqnarray}
where the inner product of operators $(A,B^\dagger)$ is determined by Eq.~(\ref{product}). The operators $A_n$ obtained from the recurrence relation (\ref{Lanczos}) are orthogonal,
\begin{equation}\label{orthogonality}
\left(A_n,A_{n'}^\dagger\right)\propto\delta_{nn'}.
\end{equation}
The residual term $R({\bf  k}\omega)$ in Eq.~(\ref{cfraction}) is given by the relations
\begin{equation}\label{residual}
R({\bf  k}t)=\left(A_{2t},A_2^\dagger\right)\left(A_2,A_2^\dagger\right)^{-1},\quad
R({\bf  k}\omega)=-i\int^\infty_0 d t e^{i\omega t}R({\bf  k}t),
\end{equation}
where the time dependence of the operator $A_{2t}$ is determined by the equation
\begin{equation}\label{timedep}
i\frac{d}{d t}A_{2t}=(1-P_0)(1-P_1)[A_{2t},H],\quad A_{2,t=0}=A_2,
\end{equation}
with the projection operators $P_n$ defined by the relation $P_nQ=\left(Q,A_n^\dagger \right)\left(A_n,A_n^\dagger \right)^{-1}A_n$.

From Eq.~(\ref{Lanczos}) and the definition of the inner product (\ref{product}) it can be seen that in the continued fraction (\ref{cfraction})
\begin{equation}\label{E0E1}
E_0=E_1=0.
\end{equation}
Using these results in Eqs.~(\ref{KuboGreen}) and (\ref{cfraction}) we obtain Eq.~(\ref{GfD}). In these equations,
$$h_{\bf  k}=\left(A_1,A_1^\dagger\right)=\left\langle\left[s^z_{\bf  k},-i\dot{s}^z_{\bf  -k}\right]\right\rangle$$
with $i\dot{s}^z_{\bf  -k}=\left[s^z_{\bf  -k},H\right]$ and
\begin{equation}\label{polfrq}
\Pi({\bf  k}\omega)=V_1R({\bf  k}\omega),\quad \omega^2_{\bf  k}=V_0.
\end{equation}

To calculate Green's function (\ref{GfD}) we have to estimate quantities (\ref{polfrq}). In accordance with Eqs.~(\ref{Lanczos}) and (\ref{E0E1}), the operator $A_2$, which enters into the polarization operator $\Pi({\bf  k}\omega)$, is equal to
\begin{equation}\label{A2}
A_2=i^2\ddot{s}^z_{\bf  k}-\omega^2_{\bf  k}s^z_{\bf  k}.
\end{equation}
The quantity $\omega^2_{\bf  k}$ is given by the relation
\begin{equation}\label{w2}
\omega^2_{\bf  k}=h_{\bf  k}\left(s^z_{\bf  k},s^z_{\bf  -k}\right)^{-1}.
\end{equation}
The second term in the right-hand side of Eq.~(\ref{A2}) ensures the orthogonality (\ref{orthogonality}) of the operators $A_0=s_{\bf  k}^z$ and $A_2$. Notice that the orthogonality of the operator
\begin{equation}\label{A1}
A_1=i\dot{s}^z_{\bf  k}=\frac{1}{2\sqrt{N}}\sum_{\bf  k'}\bigg[ \sum_\sigma\left(t_{\bf  k'}-t_{\bf  k'-k}\right)a^\dagger_{\bf  k'\sigma}a_{\bf  k'-k,\sigma}
-\left(J_{\bf  k'}-J_{\bf  k'-k}\right) s^{+1}_{\bf  k'}s^{-1}_{\bf  k'-k}\bigg]
\end{equation}
to the operators $A_0$ and $A_2$ is provided by conditions (\ref{E0E1}). Thus, the calculation of $A_2$ is reduced to the separation of the quantity $i^2\ddot{s}^z_{\bf  k}$ into a part proportional to $s^z_{\bf  k}$ and a part orthogonal to this latter operator, in accord with the Mori orthogonalization procedure.\cite{Mori} This separation cannot be performed exactly, since the quantity $\left(s^z_{\bf  k},s^z_{\bf  -k}\right)$ in Eq.~(\ref{w2}) cannot be directly calculated. To carry out the separation approximately we use the decoupling, which is similar to that applied for the Heisenberg model.\cite{Kondo,Shimahara,Tserkovnikov} For the $t$-$J$ model the analogous approach with some additional approximations was used in Refs.~\onlinecite{Vladimirov,Sherman03,Eremin}.

The operator $i^2\ddot{s}^z_{\bf  k}$ reads
\begin{eqnarray}
i^2\ddot{s}^z_{\bf  k}&=&\frac{1}{N}\sum_{\bf  qq'\sigma}f_1({\bf  kqq'})\sigma s^{-\sigma}_{-\sigma\bf k'}a^\dagger_{\bf  q,-\sigma}a_{\bf  q'\sigma}
+\frac{1}{\sqrt{N}}\sum_{\bf  q\sigma}f_2({\bf  kq})\sigma a^\dagger_{\bf  k+ q,\sigma}a_{\bf  q\sigma}\nonumber\\
&&+\frac{1}{N}\sum_{\bf  qq'\sigma}f_3({\bf  kqq'})s^z_{\bf  k'}a^\dagger_{\bf  q\sigma}a_{\bf  q'\sigma}
+\frac{1}{N}\sum_{\bf  qq'}f_4({\bf  kqq'})s^z_{\bf  k'}s^{+1}_{\bf  q}s^{-1}_{\bf  q'}\nonumber\\
&&+\frac{2}{N}\sum_{\bf  qq'}t_{\bf  q}\left(t_{\bf  q+q'}-t_{\bf  q-k+q'}\right)s^z_{\bf  k-q'}X^{00}_{\bf  q'}\nonumber\\
&&+\frac{1}{N}\sum_{\bf  qq'}J_{\bf  q}\left(J_{\bf  q+q'}-J_{\bf  q-k+q'}\right)s^z_{\bf  k-q'}X^{\uparrow\uparrow}_{\bf  q'},
\label{s2}
\end{eqnarray}
where the functions $f_i$, $t_{\bf  q}$, $J_{\bf  q}$ are given by Eqs.~(\ref{fi}) and (\ref{constants}), $X^{00}_{\bf  q}=N^{-1}\sum_{\bf  l} e^{-i{\bf  kl}}X^{00}_{\bf  l}$, $X^{\uparrow\uparrow}_{\bf  q}=N^{-1}\sum_{\bf  l} e^{-i{\bf  kl}}X^{\uparrow\uparrow}_{\bf  l}$ and $X^{\sigma\sigma'}_{\bf  l}=|{\bf  l}\sigma\rangle\langle{\bf  l}\sigma'|$.

The above equation for $i^2\ddot{s}^z_{\bf  k}$ needs in some explanation. In this equation, terms containing hole operators can be expressed in two equivalent forms. In the site representation these forms read
\begin{equation}\label{s2_1}
\left(X^{\sigma\sigma}_{\bf  l}+X^{00}_{\bf  l}\right)a^\dagger_{\bf  l'\sigma}a_{\bf  l''\sigma}
\end{equation}
and
\begin{equation}\label{s2_2}
\left(\frac{1}{2}+\frac{1}{2}X^{00}_{\bf  l}+\sigma s^z_{\bf  l}\right)a^\dagger_{\bf  l'\sigma}a_{\bf  l''\sigma}.
\end{equation}
Their equivalence follows from the completeness condition of the site states,
$$X^{00}_{\bf  l}+\sum_\sigma X^{\sigma\sigma}_{\bf  l}=1.$$
However, after the decoupling used for calculating $\Pi({\bf  k}\omega)$
this equivalence is lost. If Eq.~(\ref{s2_1}) is used, as it was done in Ref.~\onlinecite{Vladimirov}, the decoupling of the operators $X^{\sigma\sigma}_{\bf  l}$ leads to a small contribution of the decay into a fermion pair, which is assisted by charge fluctuations. In this case, the contribution of the direct decay of a spin excitation into a fermion pair is gone. As follows from Eq.~(\ref{s2_2}), this process exists. Among terms in the polarization operator this is the simplest process, which makes the main contribution in the most interesting range of frequencies $\omega<0.3t=150$~meV. As mentioned above, the incommensurability below the resonance frequency stems from this term. Besides, neglecting the decay into a fermion pair is in contradiction with the spin-wave and itinerant-electron approximations for the susceptibility, in which only this process is usually considered. Thus, representation (\ref{s2_1}) is not applicable to the considered case. Equation~(\ref{s2}) is derived using (\ref{s2_2}). This expression allows one to take proper account of the decay into a fermion pair, gives simpler decoupling and does not introduce additional Green's functions.

In all terms of Eq.~(\ref{s2}) operators belong to different lattice sites. The part of the operator $i^2\ddot{s}^z_{\bf  k}$, which is proportional to $s^z_{\bf  k}$, is contained in the last four terms in the right-hand side of this equation. This part is obtained by substituting operator multipliers of $s^z_{\bf  k}$ in these terms by their mean values. From these mean values the quantity $\omega^2_{\bf  k}$, Eq.~(\ref{wk}), is formed. From Eqs.~(\ref{a1a1}), (\ref{wk}) and (\ref{w2}) the quantity $\left(s^z_{\bf  k},s^z_{\bf  -k}\right)$ can be derived,
$$\left(s^z_{\bf  k},s^z_{\bf  -k}\right)=\frac{J|C_1|-tF_1}{4J (J\alpha|C_1|-tF_1)(\delta+1+\gamma_{\bf  k})}.$$
In the absence of holes this equation, (\ref{wk}) and (\ref{Delta}) reduce to the formulae obtained earlier for the Heisenberg antiferromagnet.\cite{Kondo,Shimahara}

The operator $A_2$ is obtained from $i^2\ddot{s}^z_{\bf  k}$ by subtracting terms entering into $\omega_{\bf  k}^2s^z_{\bf  k}$. The contribution of the two last terms in Eq.~(\ref{s2}) into $A_2$ can be neglected, since charge fluctuations are small in the $t$-$J$ model for small hole concentrations. Thus, $A_2$ contains two first terms of Eq.~(\ref{s2}) together with the third and fourth terms of this equation, in which $a^\dagger_{\bf  q\sigma}a_{\bf  q'\sigma}$ and $s^{+1}_{\bf  q}s^{-1}_{\bf  q'}$ are replaced by $a^\dagger_{\bf  q\sigma}a_{\bf  q'\sigma}-\langle a^\dagger_{\bf  q\sigma}a_{\bf  q'\sigma}\rangle$ and $s^{+1}_{\bf  q}s^{-1}_{\bf  q'}-\langle s^{+1}_{\bf  q}s^{-1}_{\bf  q'}\rangle$, respectively. Substituting $A_2$ into Eqs.~(\ref{residual}), (\ref{polfrq}), approximating the operator $A_{2t}$ by $A_2(t)$ and using the decoupling, we obtain Eq.~(\ref{Pi}).

Inserting Eqs.~(\ref{B}) and (\ref{A}) in (\ref{Pi}), for $\omega>0$ and $T=0$ we get the equation used in the above calculations,
\begin{eqnarray}
{\rm Im}\Pi({\bf  k}\omega)&=&-\frac{1}{h_{\bf  k}\omega}\Bigg\{\frac{4\pi|C_1|J}{N^2} \sum_{\bf  qq'}\left[2f^2_1({\bf  kqq'})+f^2_3({\bf  kqq'})\right]\frac{1-\gamma_{\bf  k'}}{\omega_{\bf  k'}}\nonumber\\
&&\quad\times\Bigg[\frac{Z^2[(\xi_{\bf  q}-\varepsilon_{\bf  q})(\xi_{\bf  q'}+\varepsilon_{\bf  q'})+\Delta_{\bf  q}\Delta_{\bf  q'}]}{4\xi_{\bf  q}\xi_{\bf  q'}}P(\omega_{\bf  k'}+\xi_{\bf  q'}+\xi_{\bf  q}-\omega)\nonumber\\
&&\quad +\frac{ZZ'(\xi_{\bf  q}-\varepsilon_{\bf  q})}{4\xi_{\bf  q}(4t-\Delta)}\theta(\omega-\xi_{\bf  q}-\omega_{\bf  k'}-\Delta)\nonumber\\
&&\quad +\frac{ZZ'(\xi_{\bf  q'}+\varepsilon_{\bf  q'})}{4\xi_{\bf  q'}(4t-\Delta)}\theta(\omega-\xi_{\bf  q'}-\omega_{\bf  k'}-\Delta)+\left(\frac{Z'}{8t-2\Delta}\right)^2U(\omega-\omega_{\bf  k'})\Bigg]\nonumber\\
&&+\frac{32\pi(J|C_1|)^3}{N^2}\sum_{\bf  qq'}f^2_4({\bf  kqq'})\frac{(1-\gamma_{\bf  k'})(1-\gamma_{\bf  q})(1-\gamma_{\bf  q'})}{\omega_{\bf  k'}\omega_{\bf  q}\omega_{\bf  q'}}\nonumber\\
&&\quad\times P(\omega_{\bf  k'}+\omega_{\bf  q'}+\omega_{\bf  q}-\omega)\nonumber\\
&&+\frac{2\pi}{N}\sum_{\bf  q}f_2^2({\bf  kq})\Bigg[\frac{Z^2[(\xi_{\bf  k+ q}-\varepsilon_{\bf  k+q})(\xi_{\bf  q}+\varepsilon_{\bf  q})-\Delta_{\bf  k+q} \Delta_{\bf  q}]}{4\xi_{\bf  k+q}\xi_{\bf  q}}\nonumber\\
&&\quad\times P(\omega-\xi_{\bf  k+q}-\xi_{\bf  q})
+\frac{ZZ'(\xi_{\bf  k+q}-\varepsilon_{\bf  k+q})}{4\xi_{\bf  k+ q}(4t-\Delta)}\theta(\omega-\xi_{\bf  k+q}-\Delta) \nonumber\\
&&\quad +\frac{ZZ'(\xi_{\bf  q}+\varepsilon_{\bf  q})}{4\xi_{\bf  q}(4t-\Delta)}\theta(\omega-\xi_{\bf  q}-\Delta)
+\left(\frac{Z'}{8t-2\Delta}\right)^2U(\omega)\Bigg]\Bigg\}, \label{ImPi}
\end{eqnarray}
where ${\bf  k'=k-q+q'}$ and
$$U(\omega)=\left\{\begin{array}{cl}
            0 & \mbox{if $\omega<2\Delta$,}\\
            \omega-2\Delta & \mbox{if $2\Delta<\omega<2t-\varepsilon_{\bf  Q}+\Delta$}
\end{array}\right.$$
in the considered range of frequencies.


\section*{References}
\begin{thebibliography}{99}
\bibitem{Yoshizawa}H.~Yoshizawa, S.~Mitsuda, H.~Kitazawa, and K.~Katsumata, {\it J.\ Phys.\ Soc.\ Japan} {\bf 57}, 3686 (1988).
\bibitem{Birgeneau}R.~J.~Birgeneau, Y.~Endoh, Y.~Hidaka, K.~Kakurai, M.~A.~Kastner, T.~Murakami, G.~Shirane, T.~R.~Thurston, and K.~Yamada, {\it Phys.\ Rev.} {\bf B39}, 2868 (1989).
\bibitem{Yamada}K.~Yamada, C.~H.~Lee, K.~Kurahashi, J.~Wada, S.~Wakimoto, S.~Ueki, H.~Kimura, Y.~Endoh, S.~Hosoya, G.~Shirane, R.~J.~Birgeneau, M.~Greven, M.~A.~Kastner, and Y.~J.~Kim, {\it Phys.\ Rev.} {\bf B57}, 6165 (1998).
\bibitem{Mason}T.~E.~Mason, G.~Aeppli, S.~M.~Hayden, A.~P.~Ramirez, and H.~A.~Mook, {\it Phys.\ Rev.\ Lett.} {\bf 71}, 919 (1993).
\bibitem{Matsuda}M.~Matsuda, K.~Yamada, Y.~Endoh, T.~R.~Thurston, G.~Shirane, R.~J.~Birgeneau, M.~A.~Kastner, I.~Tanaka, and H.~ Kojima, {\it Phys.\ Rev.} {\bf B49}, 6958 (1994).
\bibitem{Dai}P.~Dai, H.~A.~Mook, and F.~Do\v{g}an, {\it Phys.\ Rev.\ Lett.} {\bf 80}, 1738 (1998).
\bibitem{Arai}M.~Arai, T.~Nishijima, Y.~Endoh, T.~Egami, S.~Tajima, K.~Tamimoto, Y.~Shiohara, M.~Takahashi, A.~Garrett, and S.~M.~Bennington, {\it Phys.\ Rev.\ Lett.} {\bf 83}, 608 (1999).
\bibitem{Bourges}P.~Bourges, Y.~Sidis, H.~F.~Fong, L.~P.~Regnault, J.~Bossy, A.~Ivanov, and B.~Keimer, {\it Science} {\bf 288}, 1234 (2000).
\bibitem{Bourges98}P.~Bourges, in: {\it The Gap Symmetry and Fluctuations in High Temperature Superconductors -- NATO ASI series, Physics, vol.~371}, ed.\ J.~Bok {\it et al.} (Plenum Press, New York, 1998), p.~349.
\bibitem{He}H.~He, Y.~Sidis, P.~Bourges, G.~D.~Gu, A.~Ivanov, N.~Koshizuka, B.~Liang, C.~T.~Lin, L.~P.~Regnault, E.~Schoenherr, and B.~Keimer, {\it Phys.\ Rev.\ Lett.} {\bf 86}, 1610 (2001).
\bibitem{Haiden}S.~M.~Haiden, H.~A.~Mook, P.~Dai, T.~G.~Perring, and F.~Doðan, {\it Nature} {\bf 429}, 531 (2004).
\bibitem{Tranquada}J.~M.~Tranquada, H.~Woo, T.~G.~Perring, H.~Goka, G.~D.~Gu, G.~Xu, M.~Fujita, and K.~Yamada, {\it Nature} {\bf 429}, 534 (2004).
\bibitem{Xu}G.~Xu, J.~M.~Tranquada, T.~G.~Perring, G.~D.~Gu, M.~Fujita, and K.~Yamada, {\it Phys.\ Rev.} {\bf B76}, 014508 (2007).
\bibitem{Endoh}Y.~Endoh, T.~Fukuda, S.~Wakimoto, M.~Arai, K.~Yamada, and S.~M.~Bennington, {\it J.\ Phys.\ Soc.\ Japan} {\bf 69}, Suppl.~B, 16 (2000).
\bibitem{Stock}A.~Stock, W.~J.~L.~Buyers, R.~A.~Cowley, P.~S.~Clegg, R.~Coldea, C.~D.~Frost, R.~Liang, D.~Peets, D.~Bonn, W.~N.~Hardy, and R.~J.~Birgeneau, {\it Phys.\ Rev.} {\bf B71}, 024522 (2005).
\bibitem{Rossat}J.~Rossat-Mignot, L.~P.~Regnault, C.~Vettier, P.~Bourges, P.~Burlet, J.~Bossy, J.~Y.~Henry, and G.~Lapertot, {\it Physica} {\bf C185-189}, 86 (1991).
\bibitem{Lavagna}M.~Lavagna and G.~Stemmann, {\it Phys.\ Rev.} {\bf B49}, 4235 (1994).
\bibitem{Liu}D.~Z.~Liu, Y.~Zha, and K.~Levin, {\it Phys.\ Rev.\ Lett.} {\bf 75}, 4130 (1995).
\bibitem{Bulut}N.~Bulut and D.~J.~Scalapino, {\it Phys.\ Rev.} {\bf B53}, 5149 (1996).
\bibitem{Brinckmann}J.~Brinckmann and P.~A.~Lee, {\it Phys.\ Rev.\ Lett.} {\bf 82}, 2915 (1999).
\bibitem{Norman}M.~R.~Norman, {\it Phys.\ Rev.} {\bf B61}, 14751 (2000).
\bibitem{Uhrig}G.~S.~Uhrig, K.~P.~Schmidt, and M.~Gr\"uninger, {\it Phys.\ Rev.\ Lett.} {\bf 93}, 267003 (2004).
\bibitem{Vojta}M.~Vojta and S.~Sachdev, {\it J.\ Phys.\ Chem.\ Solids} {\bf 67}, 11 (2006).
\bibitem{Seibold}G.~Seibold and J.~Lorenzana, {\it Phys.\ Rev.} {\bf B73}, 144515 (2006).
\bibitem{Ino}A.~Ino, C.~Kim, M.~Nakamura, T.~Yoshida, T.~Mizokawa, A.~Fujimori, Z.-X.~Shen, T.~Kakeshita, H.~Eisaki, and S.~Uchida, {\it Phys.\ Rev.} {\bf B65}, 094504 (2002).
\bibitem{Eremin05}I.~Eremin, D.~K.~Morr, A.~V.~Chubukov, K.~Bennemann, and M.~R.~Norman, {\it Phys.\ Rev.\ Lett.} {\bf 94}, 147001 (2005).
\bibitem{Abanov}A.~Abanov and A.~V.~Chubukov, {\it Phys.\ Rev.\ Lett.} {\bf 83}, 1652 (1999).
\bibitem{Kimura}H.~Kimura, K.~Hirota, H.~Matsushita, K.~Yamada, Y.~Endoh, S.-H.~Lee, C.~F.~Majkrzak, R.~Erwin, G.~Shirane, M.~Greven, Y.~S.~Lee, M.~A.~Kastner, and R.~J.~Birgeneau, {\it Phys.\ Rev.} {\bf B59}, 6517 (1999).
\bibitem{Fujita}M.~Fujita, H.~Goka, K.~Yamada, and M.~Matsuda, {\it Phys.\ Rev.\ Lett.} {\bf 88}, 167008 (2002).
\bibitem{Barisic}S.~Bari\v{s}i\'c and J.~Zelenko, {\it Solid State Commun.} {\bf 74}, 367 (1990).
\bibitem{Pickett}W.~E.~Pickett, R.~E.~Cohen, and H.~Krakauer, {\it Phys.\ Rev.\ Lett.} {\bf 67}, 228 (1991).
\bibitem{Sherman08}A.~Sherman and M.~Schreiber, {\it Phys.\ Rev.} {\bf B77}, 155117 (2008).
\bibitem{Kimura00}H.~Kimura, K.~Hirota, C.~H.~Lee, K.~Yamada, and G.~Shirane, {\it J.\ Phys.\ Soc.\ Japan} {\bf 69}, 851 (2000).
\bibitem{Sherman04}A.~Sherman and M.~Schreiber, {\it Phys.\ Rev.} {\bf B69}, 100505(R) (2004).
\bibitem{Sega}I.~Sega, P.~Prelov\v{s}ek, {\it Phys.\ Rev.} {\bf B73}, 092516 (2006).
\bibitem{Mori}H.~Mori, {\it Progr.\ Theor.\ Phys.} {\bf 34}, 399 (1965).
\bibitem{Barzykin}V.~Barzykin and D.~Pines, {\it Phys.\ Rev.} {\bf B52}, 13585 (1995).
\bibitem{Sherman03a}A.~Sherman and M.~Schreiber, {\it Phys.\ Rev.} {\bf B68}, 094519 (2003).
\bibitem{Vladimirov}A.~A.~Vladimirov, D.~Ihle, and N.~M.~Plakida, {\it Phys.\ Rev.}  {\bf B80}, 104425 (2009); {\bf B83}, 024411 (2011).
\bibitem{Abrikosov}A.~A.~Abrikosov, L.~P.~Gor'kov, and I.~E.~Dzyaloshinskii, {\it Methods of Quantum Field Theory in Statistical Physics} (Pergamon Press, New York, 1965).
\bibitem{Varma}C.~M.~Varma, P.~B.~Littlewood, S.~Schmitt-Rink, E.~Abrahams, and A.~E.~Ruckenstein, {\it Phys.\ Rev.\ Lett.} {\bf 63}, 1996 (1989).
\bibitem{Kondo}J.~Kondo and K.~Yamaji, {\it Progr.\ Theor.\ Phys.} {\bf 47}, 807 (1972).
\bibitem{Shimahara}H.~Shimahara and S.~Takada, {\it J.\ Phys.\ Soc.\ Japan} {\bf 61}, 989 (1992).
\bibitem{Sherman03}A.~Sherman and M.~Schreiber, {\it Eur.\ Phys.\ J.} {\bf B32}, 203 (2003).
\bibitem{Sherman99}A.~Sherman and M.~Schreiber, in: {\it Pseudogap in High Temperature
Superconductors -- Studies of High Temperature Superconductors, vol.~27}, ed.\ A.~V.~Narlikar (Nova Science Publishers, New York, 1999), p.~163.
\bibitem{McMahan}A.~K.~McMahan, J.~F.~Annett, and R.~M.~Martin, {\it Phys.\ Rev.} {\bf B42}, 6268 (1990).
\bibitem{Gavrichkov}V.~A.~Gavrichkov, S.~G.~Ovchinnikov, A.~A.~Borisov, and E.~G.~Goryachev, {\it JETP (Russia)} {\bf 91}, 369 (2000).
\bibitem{Radtke}R.~J.~Radtke and M.~R.~Norman, {\it Phys.\ Rev.} {\bf B50}, 9554 (1994).
\bibitem{Schabel}M.~C.~Schabel, C.-H.~Park, A.~Matsuura, Z.-X.~Shen, D.~A.~Bonn, R.~Liang, and W.~N.~Hardy, {\it Phys.\ Rev.} {\bf B57}, 6090 (1998).
\bibitem{Macridin}A.~Macridin, M.~Jarrell, T.~Maier, and G.~A.~Sawatzky, {\it Phys.\ Rev.} {\bf B71}, 134527 (2005).
\bibitem{Katano}S.~Katano, M.~Sato, K.~Yamada, T.~Suzuki, and T.~Fukase, {\it Phys.\ Rev.} {\bf B62}, R14677 (2000).
\bibitem{Lake}B.~Lake, H.~M.~R\mbox{\o}nnow, N.~B.~Christensen, G.~Aeppli, K.~Lefmann, D.~F.~McMorrow, P.~Vorderwisch, P.~Smeibidl, N.~Mangkorntong, T.~Sasagawa, M.~Nohara, H.~Takagi, and T.~E.~Mason, {\it Nature} {\bf 415}, 299 (2002).
\bibitem{Kofu}M.~Kofu, S.-H.~Lee, M.~Fujita, H.-J.~Kang, H.~Eisaki, and K.~Yamada, {\it Phys.\ Rev.\ Lett.} {\bf 102}, 047001 (2009).
\bibitem{Haug}D.~Haug, V.~Hinkov, A.~Suchaneck, D.~S.~Inosov, N.~B.~Christensen, Ch.~Niedermayer, P.~Bourges, Y.~Sidis, J.~T.~Park, A.~Ivanov, C.~T.~Lin, J.~Mesot, and B.~Keimer, {\it Phys.\ Rev.\ Lett.} {\bf 103}, 017001 (2009).
\bibitem{Shen}K.~M.~Shen, F.~Ronning, D.~H.~Lee, F.~Baumberger, N.~J.~C.~Ingle, W.~S.~Lee, W.~Meevasana, Y.~Kohsaka, M.~Azuma, M.~Takano, H.~Takagi, and Z.-X.~Shen, {\it Science} {\bf 307}, 901 (2005).
\bibitem{Kanigel}A.~Kanigel, M.~R.~Norman, M.~Randeria, U.~Chatterjee, S.~Souma, A.~Kaminski, H.~M.~Fretwell, S.~Rosenkranz, M.~Shi, T.~Sato, T.~Takahashi, Z.~Z.~Li, H.~Raffy, K.~Kadowaki, D.~Hinks, L.~Ozyuzer, and J.~C.~Campuzano, {\it Nature Physics} {\bf 2}, 447 (2006).
\bibitem{Sherman87}A.~V.~Sherman, {\it J.\ Phys.} {\bf A20}, 569 (1987); A.~Sherman and M.~Schreiber, {\it Phys.\ Rev.} {\bf B65}, 134520 (2002).
\bibitem{Cullum}J.~Cullum and R.~A.~Willoughby, {\it Lanczos algorithms for large symmetric eigenvalue computations} (Birkh\"auser, Boston, 1985).
\bibitem{Tserkovnikov}Yu.~A.~Tserkovnikov, {\it Theor.\ Math.\ Phys.} {\bf 52}, 712 (1982).
\bibitem{Eremin}M.~V.~Eremin, A.~A.~Aleev, and I.~M.~Eremin, {\it JETP Lett.} {\bf 84}, 167 (2006).
\end{thebibliography}
\end{document}